\documentclass[aps,pra,twocolumn,groupedaddress]{revtex4-2}
\usepackage{units}
\usepackage{amsmath}
\usepackage{amssymb,amsfonts}
\usepackage{graphicx}
\usepackage{bm}
\usepackage{multirow,color,relsize,microtype}
\usepackage{dcolumn}
\usepackage{xcolor}
\usepackage{lmodern}

\newcommand{\be}{\begin{equation}}
\newcommand{\ee}{\end{equation}}
\newcommand{\pt}{{$\cal PT$}}

\begin{document}

\title{Insight of the Green's function as a defect state in a boundary value problem}

\author{Jose D. H. Rivero}
\author{Li Ge}
\affiliation{College of Staten Island, CUNY, Staten Island, New York 10314, USA}
\affiliation{The Graduate Center, CUNY, New York, NY 10016, USA}
\date{\today}

\begin{abstract}
A new perspective of the Green’s function in a boundary value problem as the only eigenstate in an auxiliary formulation is introduced. In this treatment, the Green's function can be perceived as a defect state in the presence of a $\delta$-function potential, the height of which depends on the Green's function itself. This approach is illustrated in one-dimensional and two-dimensional Helmholtz equation problems, with an emphasis on systems that are open and have a non-Hermitian potential. We then draw an analogy between the Green's function obtained this way and a chiral edge state circumventing a defect in a topological lattice, which shines light on the local minimum of the Green's function at the source position.
\end{abstract}

\maketitle

\section{Introduction}

Initially distributed to just 51 private members of a subscription library in 1828 \cite{green_book}, the essay of George Green on the application of mathematical analysis to the theories of electricity and magnetism has inspired generations of physicists and mathematicians. By studying the three-dimensional Laplace equation governing the static electric potential, Green recognized the unique utility of a function inversely proportional to the distance between two charged bodies, which now bears his name and preceded Dirac's introduction of the $\delta$-function by more than a century. The importance of the Green's function, together with the Green's theorem, was recognized by masterminds of modern physics such as Lord Kelvin and Julian Schwinger.

Besides its effectiveness as a theoretical and numerical tool to solve linear differential equations \cite{Stakgold2011,Melnikov2017}, the Green's function provides a fundamental connection between the source and the field \cite{Morse1953}. When extended to address dynamical equations and their spectral representations, the Green's function becomes the propagator that is fundamental to field theories \cite{schwinger_book}. As such, it is a powerful and essential tool to study a variety of transport and scattering problems, ranging from condensed matter physics \cite{Rickayzen1980,Economou2007}, optics and photonics \cite{Davy2015,Lin2016,Pick2017} to high energy physics \cite{Newton1982}.

In optical physics, one area of interest in this article, the Green's function has been routinely employed to analyze rich optical phenomena including light-matter interaction in lasers and other light sources \cite{Tureci2006,Tureci2008,Ge2010}, as well as to devise applications for optical computing, information processing, and telecommunications. What we offer here is a novel perspective of the Green's function, i.e., treating it as the single eigenstate in an auxiliary boundary value problem. In addition to further enrichment and shaping of our physical intuition through the Green's function, we find exceptional parallels between the Green's function and defect states due to a local potential, including a chiral edge state circumventing a defect on its path in a topological lattice.

Below we introduce formally the Green's function problem and lay out the fundamentals of our approach. The Green’s function of an operator $\mathcal{L}$ in a variety of physics problems can be defined by
\begin{equation}
	[z-\mathcal{L}]G(\bm{r},\bm{r'};z)=\delta(\bm{r}-\bm{r'})\label{eq:G_general}
\end{equation}
with a proper boundary condition. Here $z$ is a parameter (e.g., the energy) and $\bm{r}, \bm{r'}$ are the coordinates of the field and the source, respectively. A new perspective of the Green's function can be obtained using the following auxiliary eigenvalue problem
\begin{equation}
	[z-\mathcal{L}]\psi_m(\bm{r},\bm{r'})=\lambda_m(\bm{r'}) f(\bm{r},\bm{r'})\psi_m(\bm{r},\bm{r'}).\label{eq:G_new_def}
\end{equation}
Below we suppress the $\bm{r'}$-dependence of $\psi_m$ and $\lambda_m$ for conciseness. As we will show, this auxiliary problem has a \textit{single} eigenvalue $\lambda_0$ and eigenstate $\psi_0(\bm{r})$, \textit{when} $f(\bm{r},\bm{r'})$ is chosen to be $\delta(\bm{r}-\bm{r'})$. The Green's function is then uniquely determined by
\begin{equation}
	G(\bm{r},\bm{r'})=\frac{\psi_0(\bm{r})}{\lambda_0\psi_0(\bm{r'})},\label{eq:G_new}
\end{equation}
and $\lambda_0^{-1}$ gives the value of the Green's function at the source $\bm{r}=\bm{r'}$. As a bonus, we obtain directly the local density of states (LDOS) that is proportional to the imaginary part of $G(\bm{r},\bm{r};z)$ \cite{Economou2007}, i.e., $\text{Im}[\lambda_0^{-1}]$. The reciprocity of the Green's function, though implicit in Equation~(\ref{eq:G_new}), can be easily verified in the absence of an effective magnetic field, as shown in Appendix \ref{sec:reciproc}].

Before we apply this approach to various Hermitian and non-Hermitian problems, we first prove Equation~(\ref{eq:G_new}) and discuss the general properties of $\psi_{m}(\bm{r})$. With the choice of $f(\bm{r},\bm{r'})$ mentioned above, the right hand side of Equation~(\ref{eq:G_new_def}) for $\psi_0(\bm{r})$ is simply
\begin{equation}
	\lambda_0\delta(\bm{r}-\bm{r'})\psi_0(\bm{r})=\lambda_0\delta(\bm{r}-\bm{r'})\psi_0(\bm{r'}),\label{eq:trick}
\end{equation}
from which Equation~(\ref{eq:G_new}) follows directly by comparing with Equation~(\ref{eq:G_general}), with the requirement $\psi_0(\bm{r'})\neq 0$. 
If there were another eigenstate $\psi_{m\neq 0}(\bm{r})$, then by repeating the same procedure we would find that the Green's function is  proportional to $\psi_{m\neq 0}(\bm{r})$ as well, which contradicts the uniqueness of the Green's function with a properly imposed boundary condition. When implemented numerically, there do exist spurious eigenvectors $\psi_{m\neq 0}(\bm{r})$, which however can be easily discarded due to their ill-behaved $\lambda_m$'s as we will discuss in Section~\ref{sec:1A}. 


The auxiliary eigenvalue approach equips us with a conceptually different way to treat the Green's function, i.e., as a defect state \cite{qi_defect_2018} emerging due to the $\delta$-function potential:
\begin{equation}
	\left[\mathcal{L}+\lambda_0\delta(\bm{r}-\bm{r'})\right]\psi_0(x)=z\psi_0(x).\label{eq:defect}
\end{equation}
As we will show, this point of view is particularly interesting and helpful in a topological system with chiral edge states \cite{ hafezi_robust_2011,hafezi_imaging_2013,bandres_topological_2018,zhao_non-hermitian_2019,barik_topological_2018}. For example, if a point source is placed at the edge of a two-dimensional (2D) topological insulator, the auxiliary eigenvalue approach indicates that an analogy exists between the Green's function and a chiral edge state circumventing a defect at the same location. This interpretation provides an intuitive understanding of the local minimum of the Green's function at the source position, which we will illustrate using a 2D square lattice with a $\pi/2$ Landau gauge.

The rest of the paper is organized into two main parts, where we validate our method and discuss the new insight it provides, respectively. In Section~\ref{sec:II}, we first validate our method in a 1D Hermitian (closed) system where the analytical form of the Green's function is available. We then extend the validation to two non-Hermitian 1D systems with parity-time (\pt) symmetry \cite{Bender1998,Feng2017}, focusing on the Green's function at an exceptional point (EP). An EP is a unique degeneracy found only in non-Hermitian systems, where two or more eigenstates of the system coalesce. It has led to a plethora of intriguing phenomena, and in particular, it was shown recently that the Green's function can be fully decoupled from the coalesced eigenstate in a photonic system, which is instead given by the Jordan vector or the ``missing dimension'' of the Hilbert space \cite{Chen2020}. We show that our method based on Equation~(\ref{eq:G_new_def}) captures this extraordinary behavior nicely in a ring cavity, besides describing correctly the Green's function in a \pt-symmetric photonic molecule. We further validate our method in quasi-1D waveguides, which are frequently employed to study disordered mesoscopic and optical systems \cite{Beenakker1997,Imry2008,Mello2004}. In Section \ref{sec:III}, we first discuss how the defect state corresponding to the Green's function is conceptually different from previous studies of defect states in a 1D continuous Hermitian system and a 1D non-Hermitian lattice. We then highlight the intriguing manifestation of the linkage between the Green's function and a defect state in the aforementioned 2D topological lattice.
\vspace{-10pt}
\section{Validation} \label{sec:II}

\subsection{1D Hermitian case}
\label{sec:1A}

\begin{figure}[b]\centering
	\includegraphics[width=\linewidth]{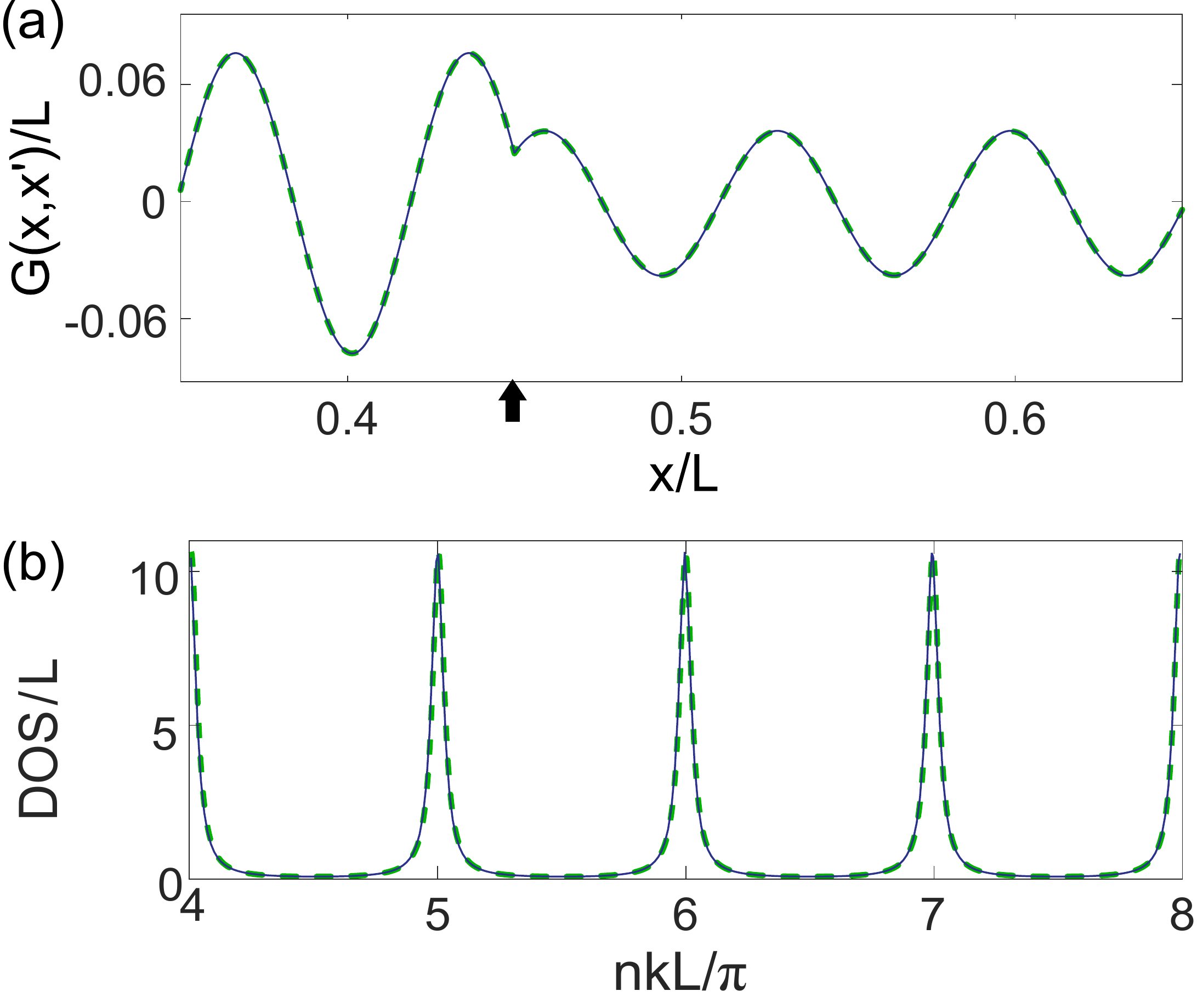}
	\caption{The Green's function (a) and DOS (b) in a 1D dielectric cavity with perfect mirrors. The green dashed lines are obtained using the analytical expression (\ref{eq11}), while the black solid lines are from our auxiliary eigenvalue approach (\ref{eq:G_new}). $n=3$, $k = 30/L$ and $x' = 0.45 L$ (marked by the arrow) are used in (a), and 2000 grid points are used for the finite difference implementation of Equation~(\ref{eq:G_new_def}).}\label{fig:1D_Hermitian}
\end{figure}

We start with an exmaple where the analytical form of the Green's function is available. Consider the scalar Helmholtz equation in 1D with a uniform refractive index $n\in\mathbb{R}$:
\begin{equation}
	\mathcal{L}=-\frac{1}{n^2}\partial_x^2,\quad z = k^2.\label{eq:Helmholtz}
\end{equation}
Here $k$ is the real-valued wave vector in free space. Below we take the speed of light in vacuum to be 1 and do not distinguish $k$ from the (circular) frequency. We impose the Dirichlet boundary conditions $G(x,x')=0$, $\psi_m(x)=0$ at $x=0,L$, which renders the system Hermitian. Consequently, it can be shown that the Green's function at the source position (i.e., $\lambda_0^{-1}$) is real:
\be
\lambda_0=\frac{k}{n}\left[\frac{1}{\tan[nk(x'-L)]}-\frac{1}{\tan(nkx')}\right].\label{eq:lambda_1D_Hermitian}
\ee
When solved using a finite difference scheme \cite{Ge2017}, the numerical value of $\lambda_0$ given by Equation~(\ref{eq:G_new}) shows a good agreement with the analytical result given by Equation~(\ref{eq:lambda_1D_Hermitian}). The corresponding Green's function, which we shown in Figure \ref{fig:1D_Hermitian}(a), also agrees nicely with its analytical expression
\begin{equation}
	G(x,x')=\begin{cases}
		\frac{\sin[nk(x-L)]}{\lambda_0\sin[nk(x'-L)]}, & (x>x')\\
		\frac{\sin(nkx)}{\lambda_0\sin(nkx')}. & (x\leq x')
	\end{cases}\label{eq11}
\end{equation}



The numerical implementation of the $\delta$-function is usually taken to be the limit of a sharp analytical distribution, such as a Gaussian with the standard deviation $\sigma \rightarrow 0$. At the same time, another small quantity that requires attention in the finite difference implementation of Equation~(\ref{eq:G_new_def}) is the lattice spacing $\Delta x$. If we choose $\sigma>\Delta x$, we find spurious eigenstates $\psi_{m\neq0}(x)$ with almost identical spatial dependence away from the source position (not shown), but they have structures that reside in the finite extension of the approximated $\delta$-function (e.g., fast oscillations). This problem is remedied by letting $\sigma<\Delta x$, which practically leads to an approximation of the $\delta$-function that has a single non-zero element at the position of the source, the value of which is given by $1/\Delta x$. This choice warrants that the integration of the $\delta$-function is 1 over any range enclosing the source and 0 otherwise.
With this choice, we find that all spurious eigenvalues of Equation~(\ref{eq:G_new_def}) approaches infinity (e.g., $|\lambda_{m\neq0}L|>10^{17}$ in the case shown in Figure~\ref{fig:1D_Hermitian}), and the sole, physical one $\lambda_0$ is easily obtained by setting the numerical routine to search for the eigenvalue with the smallest modulus.

As mentioned in the introduction, our auxiliary eigenvalue approach also produces the LDOS directly. For a Hermitian system, a limiting procedure is needed to regulate the singularity at real-valued resonant frequencies:
\be
\text{LDOS}(x;k) = \lim_{s\to 0^+}-\frac{2k}{\pi}\text{Im} [G(x,x;k+is)],\label{eq:LDOS}
\ee
i.e., a small positive imaginary part is added to the frequency here, constructing in this way the retarded Green's function [See Appendix \ref{sec:reciproc}]. Equation (\ref{eq:LDOS}) also applies to non-Hermitian cases, as long as the complex resonant frequencies are on or below the real axis. The integration of LDOS$(x;k)$ over the whole system then gives the density of states (DOS) as a function of the frequency.

To calculate DOS using our approach based on Equation~(\ref{eq:G_new_def}), we choose a small $s=0.03/L$ and calculate $-\frac{2k}{\pi}\text{Im} [\lambda_0^{-1}]$ numerically. The result agrees well with the analytical result [see Figure~\ref{fig:1D_Hermitian}(b)], where $\lambda_0$ given by Equation~(\ref{eq:lambda_1D_Hermitian}) is used. The latter leads to
\be
\text{DOS}(k)=\sum_{m}\delta\left(k-\frac{m\pi}{nL}\right)
\ee
as expected once $s\to0$, where $k_m = m\pi/nL\,(m=1,2\ldots)$ are the real-valued resonant frequencies.

\subsection{1D non-Hermitian cases}
\label{sec:nonHermitian}

For our next validation, we study the Green's function at an EP in a photonic molecule \cite{Liertzer2012}. Such a case presents a serious challenge to the standard approach based on the eigenvalues of $\cal L$, i.e., the bilinear expansion

\begin{eqnarray}
	G(\bm{r},\bm{r'};z)=\sum_{m}&&\frac{\overline{\phi}_m(\bm{r})\phi_m(\bm{r'})}{(z-z_m) (m,m)},\label{eq:G_expansion}\\
	\mathcal{L}\phi_m(\bm{r})&&=z_m\phi_m(\bm{r}).
\end{eqnarray}
Depending on the symmetry of $\mathcal{L}$, the partner function $\overline{\phi}_m(\bm{r})$ may or may not be the same as $\phi_m^*(\bm{r})$, and $(m,n)$ is the resulting inner product of $\overline{\phi}_m(\bm{r})$ and $\phi_n(\bm{r})$. At an EP, the inner product $(m,m)$ becomes zero due to the coalescence of two or more eigenstates of the system \cite{Heiss2004,Berry2004,Feng2017}. Although this divergence can be eliminated using a non-Hermitian perturbation theory \cite{Pick2017,Chen2020}, it requires \textit{a priori} knowledge of the EPs and the generalized eigenvectors, which adds to the complexity of the problem. Our auxiliary eigenvalue formulation, on the other hand, does not suffer from this drawback.

\begin{figure}[t]
	\centering
	\includegraphics[width=\linewidth]{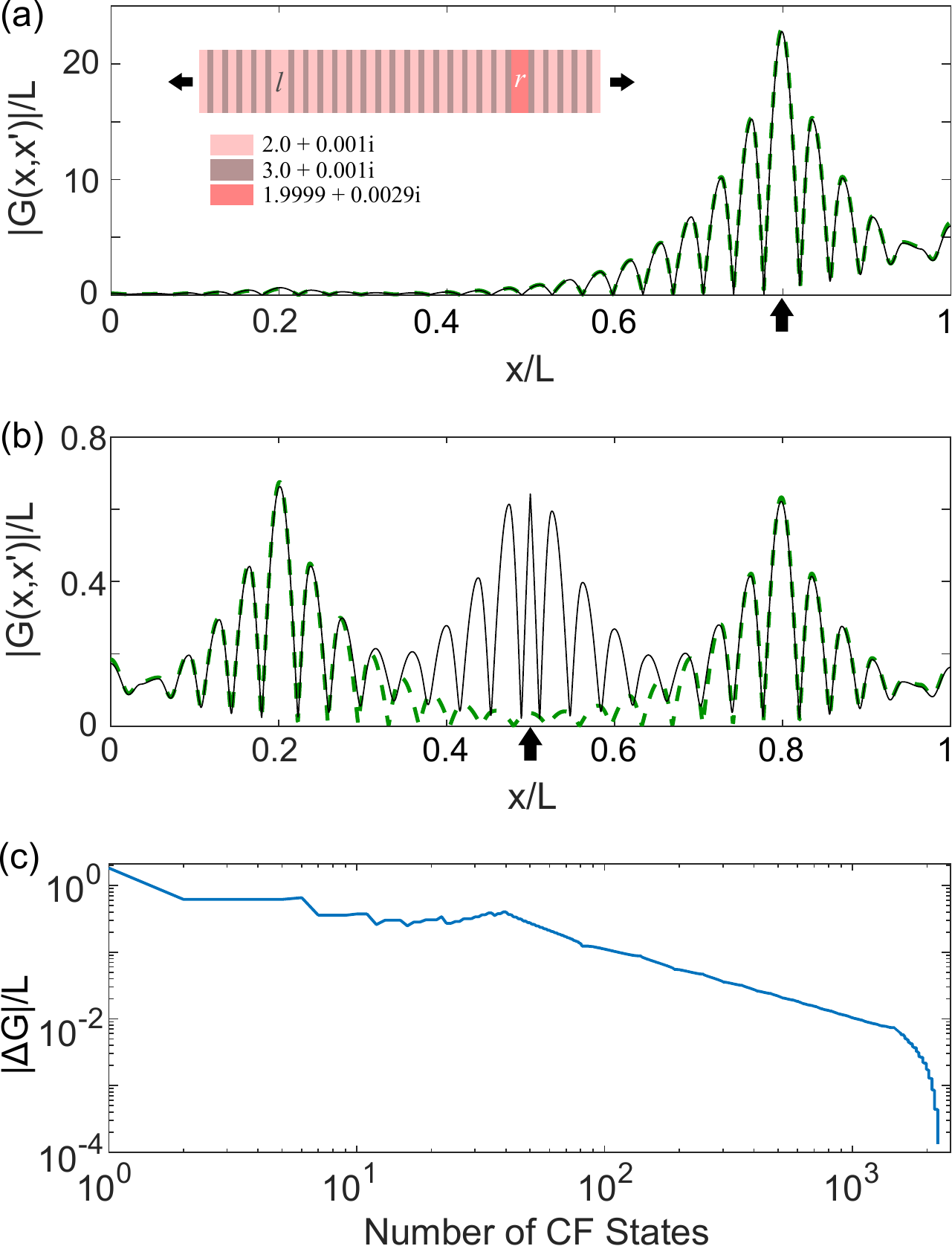}
	\caption{Green's function of a $\mathcal{PT}$ photonic molecule at an EP. The source is placed at the center of the right cavity in (a) and in between the two cavities in (b). Inset in (a): Schematic of the photonic molecule and the imposed boundary condition, and labels denoting the refractive indices of each component of the heterostructure. Green dashed lines correspond to the perturbative expression in Equation~(\ref{eq:G_EP}) and black solid lines correspond to our auxiliary eigenvalue approach. The locations of the sources are marked with arrows. (c) Difference between the bilinear expansion with the perturbative correction and the auxiliary method in (b) at the source position.
	}
	\label{fig:1D_EP}
\end{figure}

Below we exemplify our method and compare it to the result of the perturbation theory, without the need to invoke the Jordan vector that completes the Hilbert space at the EP \cite{Chen2020}. Our photonic molecule is composed of two half-wavelength cavities coupled by a distributed Bragg reflector (DBR) placed in air [See Figure \ref{fig:1D_EP}(a); inset]. If an unbalanced loss distribution is introduced in the two half-wavelength cavities, an emerging effective \pt~symmetry governs the system \cite{Ge2016}.

Here we consider the outgoing boundary condition at a real-valued frequency, which corresponds to the laser and is different from quasi-bound states or resonances with a complex frequency throughout the entire space \cite{Tureci2006}. The resulting complex eigenvalues of $\mathcal{L}$ given in Equation~(\ref{eq:Helmholtz}) are found in the lower half of the complex frequency plane, as a result of the non-Hermiticity caused by the cavity openness. These complex eigenvalues are known as continuous flux (CF) states \cite{Tureci2006,Ge2017}, which have also been used to study nuclear decays \cite{Kapur1937,Goldberger1964}. We note that the EPs of CF states have not been studied before, unlike their counterparts of quasi-bound states or resonances.

In the vicinity of an EP of frequency $k_0$, the CF states in our system can be expressed in terms of waves confined in the left- and right-cavity, i.e., $\psi(x)\approx a_l\psi_l(x)+a_r\psi_r(x)$, where the amplitudes $a_{l,r}$ are determined by the Helmholtz equation. Without the $\mathcal{PT}$-symmetric perturbation, our photonic molecule is symmetric and hence $a_l=\pm a_r$ in the symmetric and anti-symmetric modes with CF frequencies $\tilde{k}_S,\tilde{k}_A$. The introduction of a weak $\mathcal{PT}$-symmetric perturbation in the dielectric function couples the amplitudes $a_l$ and $a_r$, determined by their spatial overlap $C$ with the non-Hermitian perturbation, which represents the strength of gain and loss. 

The eigenfrequencies of the perturbed system are then found to be $q_\pm^2=k_0^2 \left(1\pm \Delta\varsigma\right)$. Here $k_0$ is the CF frequency of a single half-wavelength cavity sandwiched by two DBRs and $\Delta$ is a dimensionless detuning defined by $(\tilde{k}_S^2-\tilde{k}_A^2)/2k_0^2$. We have also defined $\varsigma\equiv\sqrt{1-\beta^2}$, where $\beta=C/\Delta$.
One EP is reached when $\varsigma=0$, resulting in $q^2_\pm=q_0^2$.  The corresponding eigenstates $\psi_\pm(x)$ are given by
\begin{equation}
	\psi_\pm(x)=\psi_l(x)+\beta_\pm\psi_r(x),
\end{equation}
where $\beta_\pm\equiv i\beta\pm\varsigma$. The inner product of the eigenstates is defined as
\be
(i,j)\equiv\int \epsilon_0(x) \psi_{i}(x)\psi_{j}(x) dx, \quad (i,j=\pm)\label{eq:inner_EP}
\ee
where $\epsilon_0(x)$ is the dielectric function before the \pt-symmetric perturbation is introduced. In other words, the partner functions in Equation~(\ref{eq:G_expansion}) are chosen as $\overline{\psi}_\pm(x)=\epsilon_0(x)\psi_\pm(x)$, similar to the Hermitian case discussed earlier. This definition of the inner product warrants the biorthogonality $(+,-)=(-,+)=0$, and we find $(+,+)=2\beta_+\varsigma\to0$, $(-,-)=-2\beta_-\varsigma\to0$ as the system approaches the EP.

As seen from Equation~(\ref{eq:G_expansion}), the vanishing inner products $(+,+),\,(-,-)$ here at the EP cause a catastrophe in the calculation of the Green's function, because the two corresponding terms diverge independent of the frequency:
\be
G(x,x';k)\approx\frac{\overline{\psi}_+(x')\psi_+(x)}{(q^2-q_+^2)(+,+)}+\frac{\overline{\psi}_-(x')\psi_-(x)}{(q^2-q_-^2)(-,-)}.\label{eq:G_EP0}
\ee
However, it can be shown that the diverging behaviors in the two terms cancel each other precisely \cite{Pick2017,Chen2020}, leading to
\begin{align}
	G(x,x';k) \approx &\;\epsilon_0(x')\frac{\psi_l(x')\psi_l(x)+\psi_r(x')\psi_r(x)}{k^2-k_0^2} \nonumber\\
	&-i\Delta k_0^2\epsilon_0(x')\frac{\psi_{EP}(x')\psi_{EP}(x)}{(k^2-k_0^2)^2}
	\label{eq:G_EP}
\end{align}
to the leading order of the small perturbation parameter $\varsigma$. Here $\psi_{EP}=\psi_\pm|_{\beta=1}$ is the coalesced eigenstate at the EP. As expected \cite{Lin2016,Heiss2015}, a second-order pole appears in the second term due to this coalescence. Details on the perturbative analysis can be found in Appendix \ref{sec:pert}.

To verify the robustness of our method based on Equation~(\ref{eq:G_new}) in the vicinity of the EP, we choose the heterostructure of length $L$ shown in Figure~\ref{fig:1D_EP}. It consists of DBRs of refractive indexes $n_1=2+0.001i$ and $n_2=3+0.001i$, and each layer accommodates a quarter of their respective wavelengths at $ka=1.570$, where $a$ is the lattice constant. 
The two half-wavelength cavities with loss are fine-tuned to achieve an EP at $k_\text{EP}a=1.570-0.006913i$ 
using $n_l=2+0.001i$,$n_r=1.9999+0.0029i$, and $k=\text{Re}[k_\text{EP}]$. We note that the loss in the right cavity is stronger than that in the left cavity, creating an effective $\mathcal{PT}$-symmetric system with its complex eigenvalue spectrum shifted along the imaginary axis \cite{Guo}.

Figure~\ref{fig:1D_EP} shows the Green's function of this system when the source is placed at different locations. First, we place the source at the center of the right cavity, where we expect the approximation (\ref{eq:G_EP}) using just $\psi_\pm(x)$ [and $\psi_{l,r}(x)$] to hold. As Figure~\ref{fig:1D_EP}(a) shows, it is nearly identical to the result of our auxiliary eigenvalue approach (\ref{eq:G_new}), and the inclusion of more CF states away from the EP barely changes the Green's function (not shown).
If instead, we place the source at the center of the heterostructure, Equation~(\ref{eq:G_EP}) alone is insufficient to capture the features of the Green's function [see Figure~\ref{fig:1D_EP}(b)]. However, once a large set of additional eigenfunctions of $\mathcal{L}$ are also included, a good agreement to our auxiliary eigenvalue approach is again observed [see Figure~\ref{fig:1D_EP}(c)]. 
These results show that Equation~(\ref{eq:G_new}) provides a reliable and convenient method to calculate the Green's function, even in the presence of EPs.

\begin{figure}[t]
	\centering
	\includegraphics[width=\linewidth]{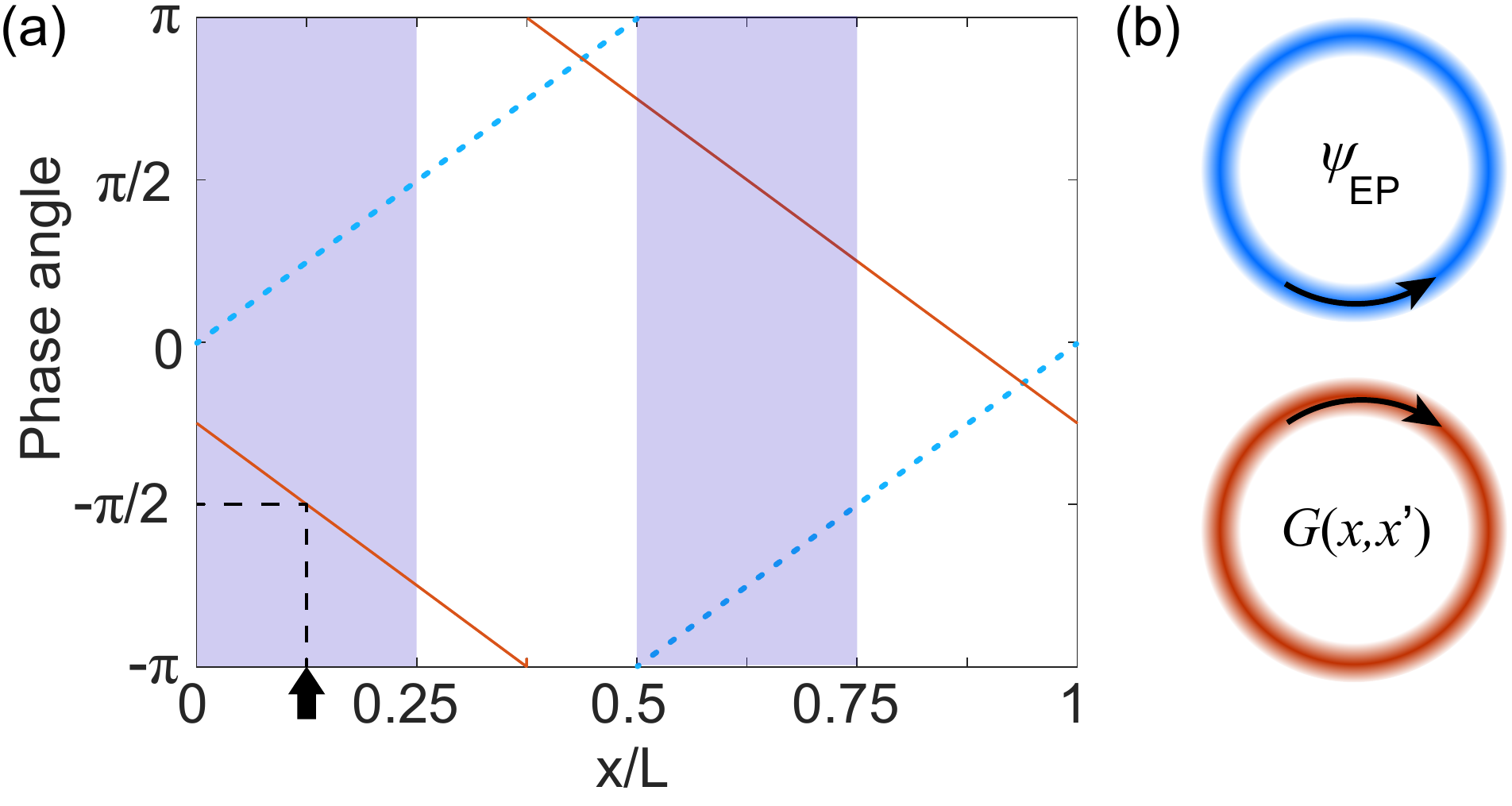}
	\caption{Chirality-reversal Green's function in a \pt-symmetric ring cavity. (a) Phase of the Green's function (solid line) and the coalesced eigenstate (dotted line). Dashed lines point to the phase of the Green's function at the source (marked by the arrow). Shaded regions show the half periods with higher loss. (b) False color plots showing the constant amplitudes of the coalesced eigenstate (top) and the Green's function (bottom). A finite width is imposed on the ring for visual clarity. Here $n_0=3$, $\delta n=0.003$, $l=1$, $k_\text{EP}L=2.0944 - 0.0021i$, $kL=2.0944$, and $x'=L/8$. }\label{fig:Jordan}
\end{figure}

A similar but more striking behavior of the Green's function at an EP was recently reported in an effective \pt-symmetric ring cavity with refractive index \cite{Chen2020}
\be
n(\theta) = (n_0 +i\delta n) + \delta n(\cos2l\theta+i\sin2l\theta).\label{eq:n_PT_ring}
\ee
Here $\theta$ is the azimuthal angle, and below we will use the arc length $x=R\theta\in[0,L]$ as the coordinate, where $R$ is the radius of the ring and $L=2\pi R$ is the circumference. $n_0$ and $\delta n$ are the real and imaginary parts of the background index, and the latter includes both absorption and radiation losses and is positive. $l$ is a positive integer, and the complex index grating, proportional to $e^{2il\theta}$, scatters the clockwise (CW) wave of angular momentum $-l$ to the counterclockwise (CCW) wave of angular momentum $l$ 
but \textit{not} vice versa. Note that the chirality and the sign of $\text{Im}[n]$ here are defined with respect to the temporal dependence $e^{-i\omega t}$.

Consequently, an EP appears at $k_\text{EP}R = l/(n_0+i\delta n)$ with the coalesced CCW eigenstate $\psi(x)=e^{il\theta}$ \cite{Miao2016}. The Green's function, on the other hand, can be fully decoupled from this mode even on resonance, if the source is placed at $\theta=(m+1/4)\pi/l$ where $m$ is a non-negative integer \cite{Chen2020}, i.e., at one of the most lossy spots in the passive ring cavity. The Green's function is given by the corresponding Jordan vector $J(x)\propto e^{-il\theta}$ instead, i.e., the ``missing dimension'' of the Hilbert space at the EP in the CW direction.

This extraordinary behavior is captured nicely using our auxiliary eigenvalue approach (see Figure~\ref{fig:Jordan}). In addition, a perturbative approach shows that the Green's function at the source is given by \cite{Chen2020}
\be 
G(x',x';k) \approx \frac{1}{2(k-k_\text{EP})k_\text{EP}L},
\ee
which is almost imaginary on resonance (i.e., $k=\text{Re}[k_\text{EP}]$) for a high-Q resonance with $|\text{Im}[k_\text{EP}]|\ll\text{Re}[k_\text{EP}]$. In the case shown in Figure~\ref{fig:Jordan}, this value is $(-113.99i + 0.228)R$ and nearly identical to that given by our auxiliary eigenvalue approach, i.e., $(-113.99i + 0.224)R$. 

\subsection{Quasi-1D waveguides}
\label{sec3}

Quasi-1D waveguides are frequently used in the study of disordered mesoscopic and optical systems \cite{Beenakker1997,Imry2008,Mello2004}, and the Green's function plays a crucial role to construct the scattering and transfer matrices \cite{fisher_relation_1981,stone_what_1988}. Here again we consider the scalar Helmholtz equation in a waveguide with background refractive index $n(\bm{r})$:
\begin{equation}
	\mathcal{L}=-\frac{1}{n^2(\bm{r})}(\partial_x^2+\partial_y^2),\quad z = k^2.\label{eq:Helmholtz2D}
\end{equation}
A finite width $L_y$ in the transverse direction and the Dirichlet boundary conditions at $y=0,L_y$ lead to a set of transverse modes (``channels'') $f_m(y)=\sin [nk_m^{(y)}y]$, where $nk_m^{(y)}=m\pi/L_y$ is the transverse wave number and $m$ is a positive integer. At a given frequency $k$, the longitudinal wave number in the $m$th channel is given by $nk_m^{(x)}=n\{k^2-[k_m^{(y)}]^2\}^{1/2}$, and this channel is propagating (evanescent) if $k_m^{(x)}$ is real (imaginary).

\begin{figure}[b]\centering
	\includegraphics[width=\linewidth]{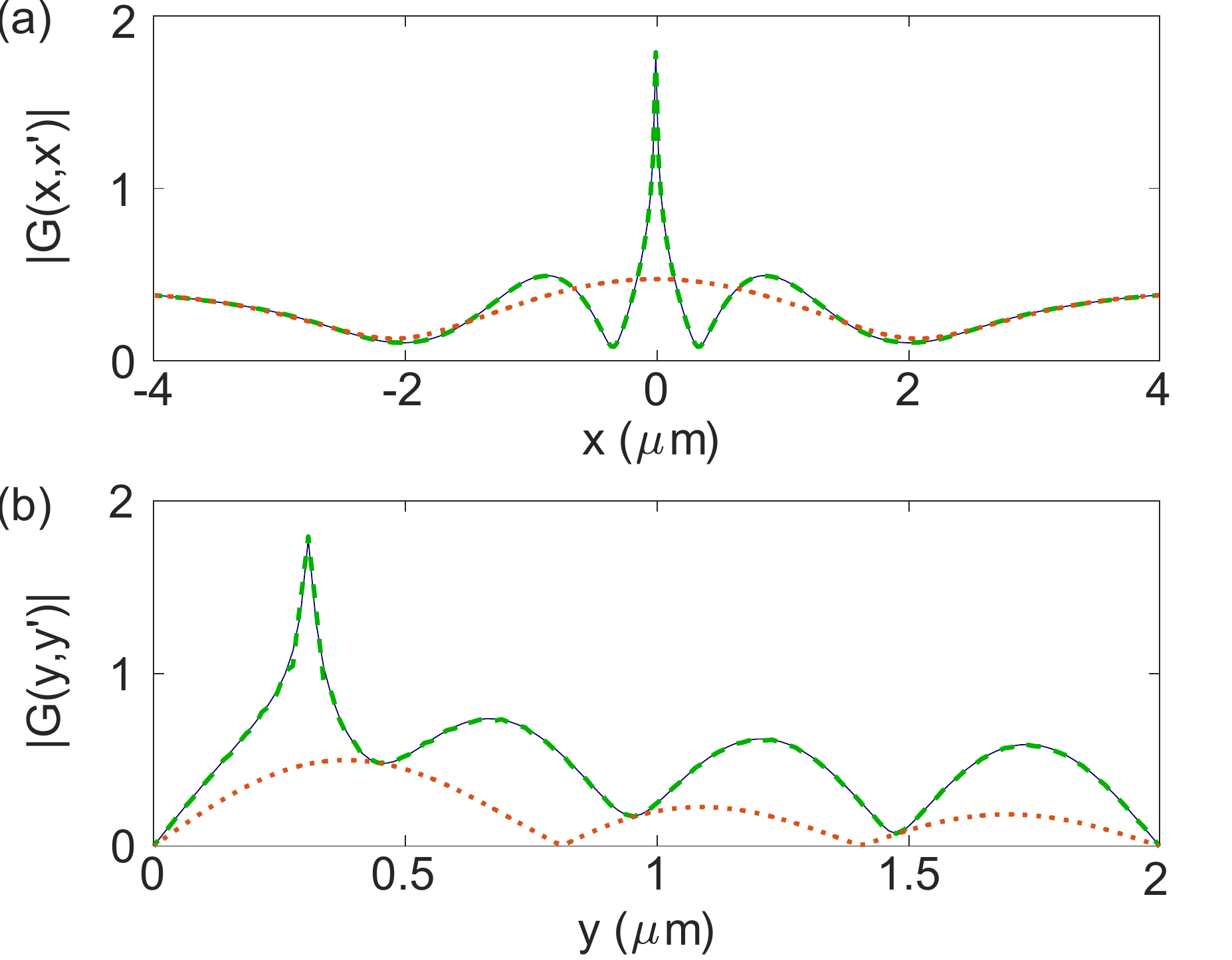}
	\caption{Green's function in a uniform waveguide with $L_y=2 \,\mu$m and $n=1.5$ everywhere. (a) and (b) show the slices along $x$ and $y$ at the source position $x'=0,\,y'=3/13\,\mu$m. The free-space wavelength is $1550$ nm. }\label{fig:WG_empty}
\end{figure}

To validate our auxiliary eigenvalue approach in quasi-1D waveguides, we first consider a uniform waveguide with the outgoing boundary condition. In this case an analytical expression exists for the Green's function, which can be written as the following infinite sum using the channel functions:
\be
G(\bm{r},\bm{r'};k) = \sum_m n\frac{\sin[nk_m^{(y)}y']\sin[nk_m^{(y)}y]}{ik_m^{(x)}L_y}e^{\pm ink_m^{(x)}x}.\label{eq:G_wg_anal}
\ee
Here the source point $\bm{r'}=(x',y')$ is placed at $x'=0$, $y'\neq0,L_y$, and the ``$+\,(-)$" sign in the exponent applies to a positive (negative) $x$. Clearly, only the propagating channels of a finite number affect the farfield behavior of the Green's function, while the logarithmic divergence of the Green's function at the source, a generic property in 2D (including quasi-1D), is reflected by the infinite number of evanescent channels in the summation. Numerically, this divergence is truncated either by the inclusion of only a finite number of evanescent channels or the finite resolution of the spatial discretization. We also note that the Green's function is dimensionless in quasi-1D.

Figure~\ref{fig:WG_empty} shows one example where our auxiliary eigenvalue approach, implemented using the finite difference method \cite{Ge_scattering_2015,Ge2017} (solid line), is compared with the analytical result given by Equation~(\ref{eq:G_wg_anal}). With just the three propagating channels available in this case (dotted line), Equation~(\ref{eq:G_wg_anal}) describes the Green's function well far from the source ($>2\,\mu{m}$); with more channels included (e.g., 100; dashed line), a good agreement
between Equation~(\ref{eq:G_wg_anal}) and our auxiliary eigenvalue approach is observed, where the grid spacings $\Delta x=1/30\,\mu$m, $\Delta y=3/130\,\mu$m used in the finite difference method are comparable to the shortest evanescent tail in the summation (i.e., $1/k_{m=100}^{(x)}\approx0.01\,\mu$m.)

\begin{figure}[b]\centering
	\includegraphics[width=\linewidth]{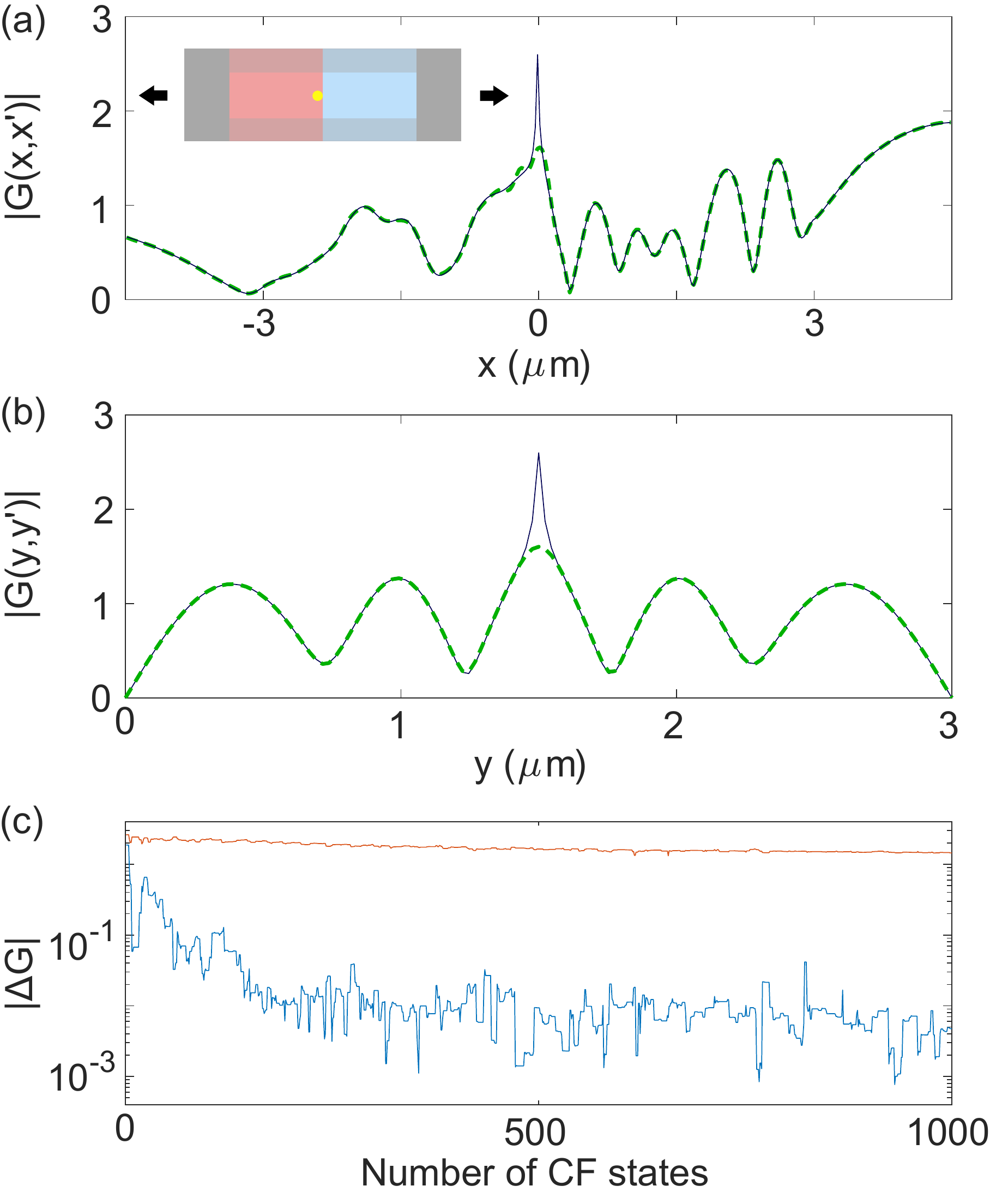}
	\caption{(a,b) Same as Figure~\ref{fig:WG_empty} but for the Green's function in a \pt-symmetric waveguide. Here $L_x=9 \,\mu$m, $L_y=3 \,\mu$m, and the source is at $x'=-1/60\,\mu$m, $y'=1.5\,\mu$m. Inset in (a): Schematic of the waveguide. $\text{Im}[n(x)]=\pm0.05$ for $x\in[-3,0]\mu$m and $[0,3]\mu$m respectively, where $\text{Re}[n(x)]=2$ in the central half of the width. $\text{Re}[n(x)]=1.2$ elsewhere inside the waveguide and $1$ outside. (c) Convergence of the bilinear expansion at the source (upper) and at $x=4.5\,\mu$m, $y=1.5\, \mu$m (lower). }\label{fig:WG}
\end{figure}

For structured or disordered quasi-1D waveguides connected to two semi-infinite regions (``leads"), the Helmholtz equation is no longer separable in $x$ and $y$, and an analytical expression for the Green's function in the form of Equation~(\ref{eq:G_wg_anal}) does not exist. Nevertheless, one can still compare to the bilinear expansion (\ref{eq:G_expansion}) as we show in Figs.~\ref{fig:WG}(a) and (b) for a \pt-symmetric waveguide, despite its slow convergence [see Figure~\ref{fig:WG}(c)].

Here we also briefly review a standard technique used to calculate the Green's function in quasi-1D waveguides, i.e., the recursive Green's function method \cite{thouless_conductivity_1981,Sols_theory_1989}. The comparison of the Green's functions obtained in our proposed approach and by the recursive Green's function method will be present in Section \ref{sec:III}.

The heart of the recursive Green's function method lies in the celebrated Dyson's equation:
\begin{equation}
	\bm{G}=\bm{G}^0 + \bm{G}^0\bm{VG}.\label{eq:RGF}
\end{equation}
$\bm{G}$ is the Green's matrix expressed in the spatial basis here, i.e., its element $\bm{G}_{\underline{i,j}}$ gives the value of the Green's function at point $i$ when the source is placed at point $j$. In a quasi-1D system with $N$ segments, $\bm{V}$ represents the couplings between the $(N-1)$th and $N$th segments due to $\mathcal{L}$, and $\bm{G}^0$ is the value of $\bm{G}$ when $\bm{V}$ is taken to be zero. In other words, $\bm{G}^0$ contains the Green's matrices of two separate systems, one for the first $N-1$ segments on the left as a whole and one for the last segment on the right. This recursive procedure then starts with a single segment on the left and proceeds with more segments added from the right, one at a time.

When implemented using the finite difference method that is equivalent to a tight-binding lattice \cite{Sols_theory_1989}, the recursive Green's function gives the identical result as our auxiliary eigenvalue approach, because they solve exactly the same discretized equation. Each segment of the system is simply a single column along $y$. We note, however, these two methods serve different purposes. The auxiliary eigenvalue approach gives the values of the Green's function inside the entire waveguide for a given position of the source, which can be both in the interior of the waveguide or on its boundaries. The recursive Green's function method, on the other hand, is particularly suitable for solving the Green's function connecting the two boundary layers, with the source placed in either one of them [See Appendix \ref{sec:RGF}]. Extra steps are needed to solve the Green's function inside the waveguide, which is more intense numerically.

\section{New insights} \label{sec:III}

Having validated the proposed auxiliary eigenvalue approach to calculate the Green's function in various systems, we now turn to the new insight this method provide, i.e., viewing the Green's function as a defect state as manifested by Equation~(\ref{eq:defect}).

We start by revisiting the 1D Hermitian system described by the scalar Helmholtz equation in Figure~\ref{fig:1D_Hermitian}. The eigenstate $\psi_0(x)$ in our auxiliary eigenvalue equation (\ref{eq:G_new_def}) (and the Green's function) for a \textit{given} energy $z=k^2$ then corresponds to the wave function given by the Schr\"odinger equation inside a closed box with a $\delta$-function potential:
\begin{equation}
	\left[-\frac{1}{n^2}\partial_x^2+\lambda_0\delta(x-x')\right]\psi_0(x)=z\psi_0(x),\label{eq:Schrodinger}
\end{equation}
and the sign of $\lambda_0$ determines whether the potential is attractive or repulsive, which can be found analytically using Equation~(\ref{eq:lambda_1D_Hermitian}).

This property of $\lambda_0$ is a crucial difference between our approach and previous studies of defect states, where the defect strength (i.e., $\lambda_0$) is treated as a free parameter to find all possible $z$'s. In a nutshell, a given defect strength leads to an infinite set of wave functions with different energies $\{z_i\}$ \cite{Patil2006}, while for a given $z$ there is a single wave function of Equation~(\ref{eq:Schrodinger}) [and more generally, Equation~(\ref{eq:defect})], i.e., the Green's function, the value of which at the source gives $\lambda_0^{-1}$. Such a one-to-one mapping from $z$ to $\lambda_0$ as an eigenvalue problem has not been recognized previously to the best of our knowledge, which has prevented the association of a Green's function as a defect state in the past.

\begin{figure}[t]
	\centering
	\includegraphics[width=\linewidth]{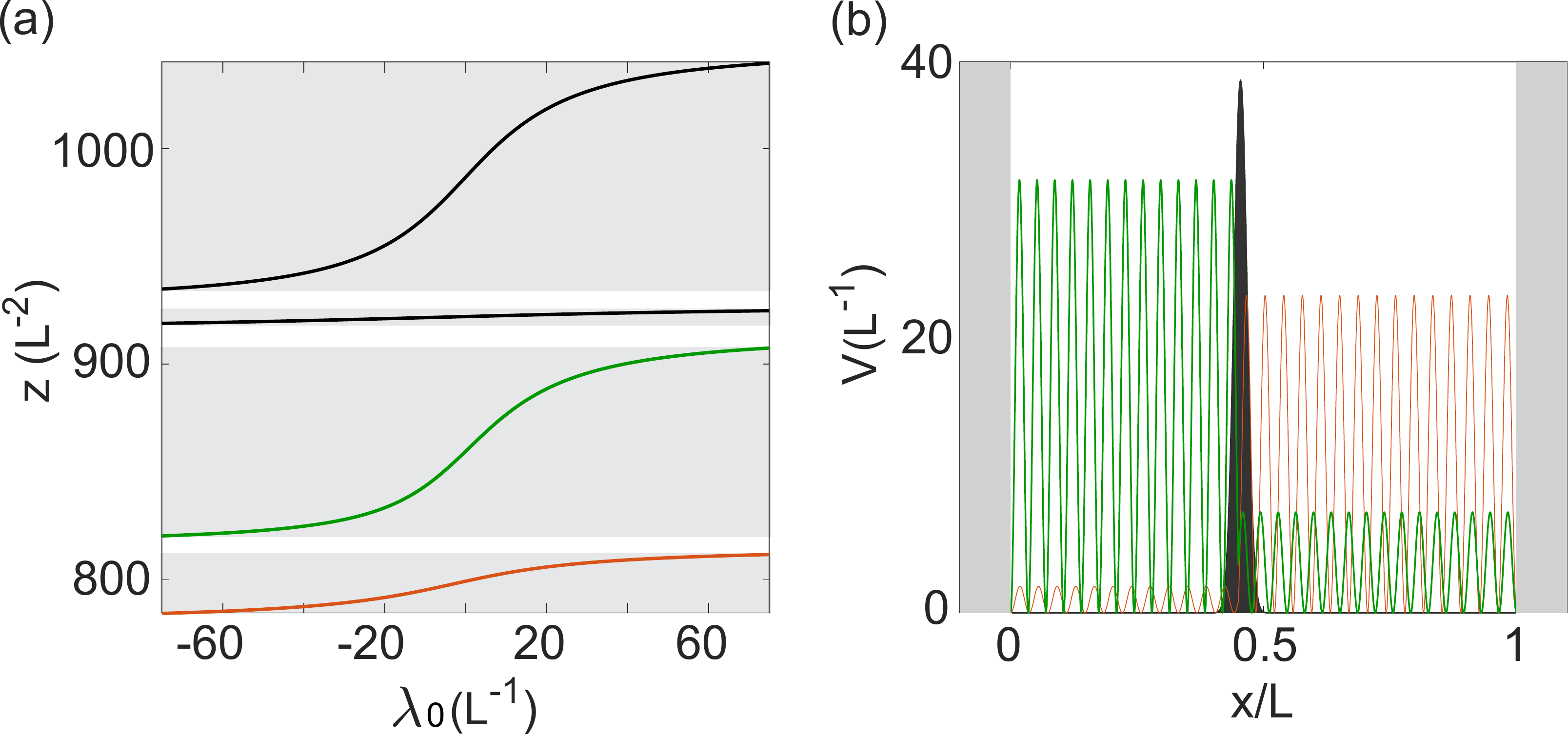}
	\caption{(a) Eigenvalues of the Schr\"odinger equation (\ref{eq:Schrodinger}) as a function of $\lambda_0$ for $n=3$ and $x' = 0.45 L$. Only four of them are shown near $z=900/L^2$. (b) Schematics showing the potential as a function of the position. Shaded areas indicate the Dirichlet boundary condition and the $\delta$-function potential with $\lambda_0=38.6/L$. Two unnormalized wave functions $|\psi(x)|^2$ are also shown.}\label{fig:Schrodinger}
\end{figure}

To illustrate this point, we show in Figure~\ref{fig:Schrodinger}(a) several energy eigenvalues $z_i$'s of Equation~(\ref{eq:Schrodinger}) as a function of  $\lambda_0$. It is important to note that the ``bandwidths'' of the eigenstates do not overlap [shaded areas in in Figure~\ref{fig:Schrodinger}(a)], even when the range of $\lambda_0$ is extended from $-\infty$ to $\infty$. This feature is determined by the uniqueness of the Green's function: for a given energy $z$ and source position $x'$, the Green's function is uniquely determined by the boundary condition, and so is its inverse height at $x'$, which gives the only allowed value of $\lambda_0$ for this pair of $z$ and $x'$. Each of the ``bandgaps'' in Figure~\ref{fig:Schrodinger}(a) shrinks to a single point when $\lambda_0$ is extended from $-\infty$ to $\infty$, and they correspond to the energy eigenvalues of the two subsystems separated by the $\delta$-potential.

A particularly interesting type of defect states has a spatially localized wave function. It takes place when the energy of the defect state is inside a bandgap, causing the defect state to be largely decoupled from the rest of the system. Here we show that the Green's function in a 1D non-Hermitian lattice can have the same property, which further stresses the new perspective our method provides.
This 1D lattice features a non-Hermitian flatband \cite{qi_defect_2018}, enabled by imposing non-Hermitian particle-hole (NHPH) symmetry \cite{Malzard,zeromodeLaser,NHC_arxiv,Kawabata} via staggered gain and loss along a lattice of optical resonators. In the tight-binding Hamiltonian description, it is given by
\be
H \phi_n = [z_0+(-1)^{n-1} i\gamma]\phi_n + g(\phi_{n+1}+\phi_{n-1}),
\ee
where $n$ labels the lattice sites from 1 to $N$, $g\in\mathbb{R}$ is the nearest neighbor coupling, and the alternate positive and negative imaginary on-site detuning $\gamma$ models gain and loss. Below we shift the energy reference such that the on-site energy $z_0$ of a resonator mode is zero. NHPH symmetry then warrants a symmetric energy spectrum about the origin of the complex plane, i.e., $z_i=-z_j^*$. At a critical value of the gain and loss strength $\gamma=2g$, the real part of the energy bands collapse to 0, with all the eigenvalues $\{z_i\}$ completely imaginary.

Now let us consider the Green's function of such a system, i.e.,
\be
(z-H)G = \delta_{nn'}.
\ee
In the auxiliary eigenvalue problem, it can be rewritten as
\be
(H+\lambda_0\delta_{nn'})\psi_n = z\psi_n,\quad G = \frac{\psi_n}{\lambda_0\psi_{n'}}.
\ee
In other words, the Green's function corresponds to a single defect state, where the on-site energy at the position of the source $n'$ is shifted by $\lambda_0$.

\begin{figure}[b]
	\centering
	\includegraphics[width=\linewidth]{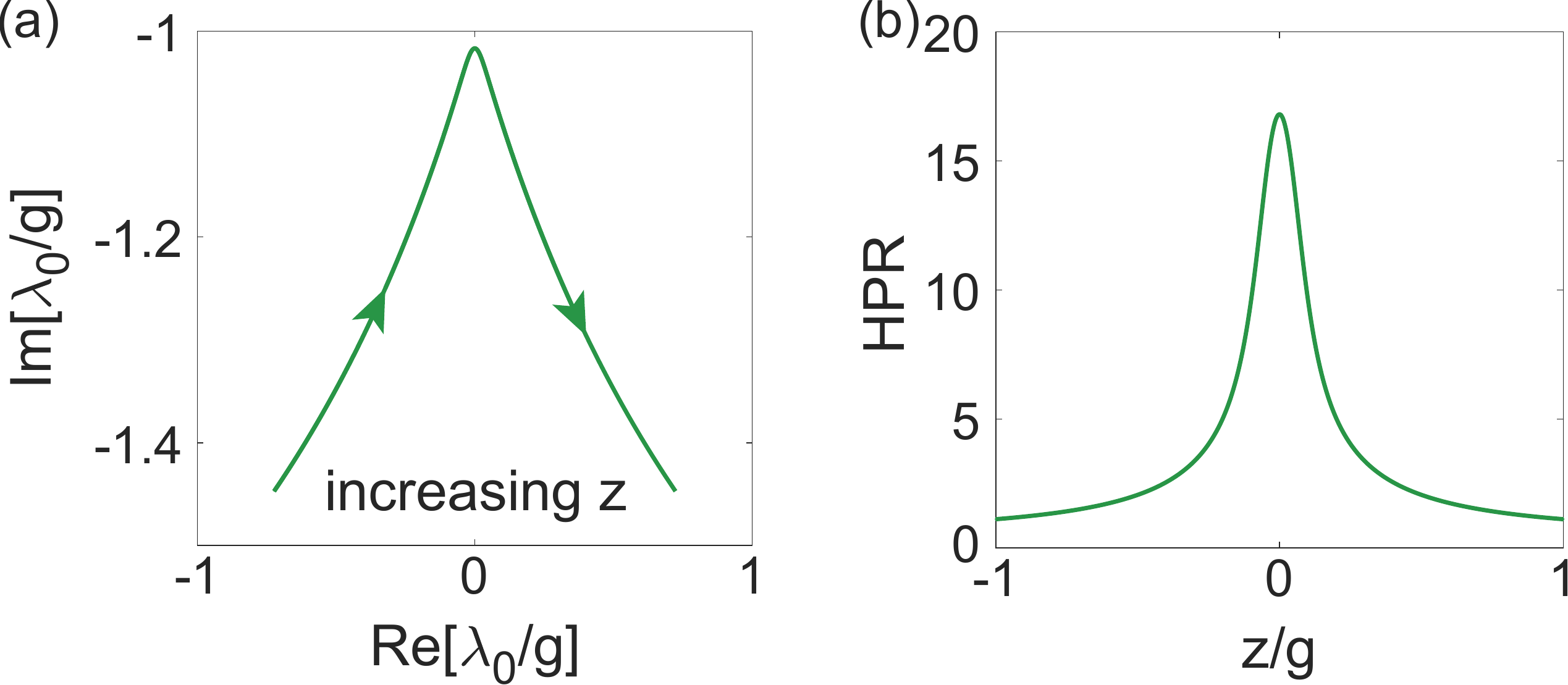}
	\caption{(a) Complex defect strength $\lambda_0$ when $z$ increases from $-g$ to $g$. The lattice has 60 sites and $\gamma=2g$. (b) HPR as a function of $z$, showing a strong localization (HPR $\sim1$) at $|z|=t$ to a weaker localization (HPR $>15$) at $z=0$.}\label{fig:NHPH}
\end{figure}

We emphasize that this shift $\lambda_0$ is \textit{complex} in general and hence different from the Hermitian case shown in Figure~\ref{fig:Schrodinger} and other typical considerations of a defect state, including that in Reference~\cite{qi_defect_2018}. One example is shown in Figure~\ref{fig:NHPH}, where we place the source at the left edge ($n'=1$) of a lattice with 60 lattice sites.  $\lambda_0$ stays in the lower complex plane when $z$ increases from $-g$ to $g$, and the stronger defect strength (and lower LDOS) at $z=\pm g$ gives the Green's function a clearly localized profile, as reflected by the half participation ratio (HPR) defined by $\sum_n \,(|\psi_n|^2)^2/(2\sum_n\,|\psi_n|^4)$ in the units of the lattice constant [see Figure~\ref{fig:NHPH}(b)]. We note that the HPR of an exponentially localized intensity $|\psi_n|^2=e^{-n/\xi}\,(n=1,2,\ldots,N)$ is approximately equal to its localization length $\xi$ \cite{Baboux} when $\xi,N\gg1$. The localization of the Green's function is weakened as the defect strength $\lambda_0$ becomes smaller near $z=0$, with its HPR now greater than 15 lattice sites.\\

\begin{figure}[b]
	\centering
	\includegraphics[width=\linewidth]{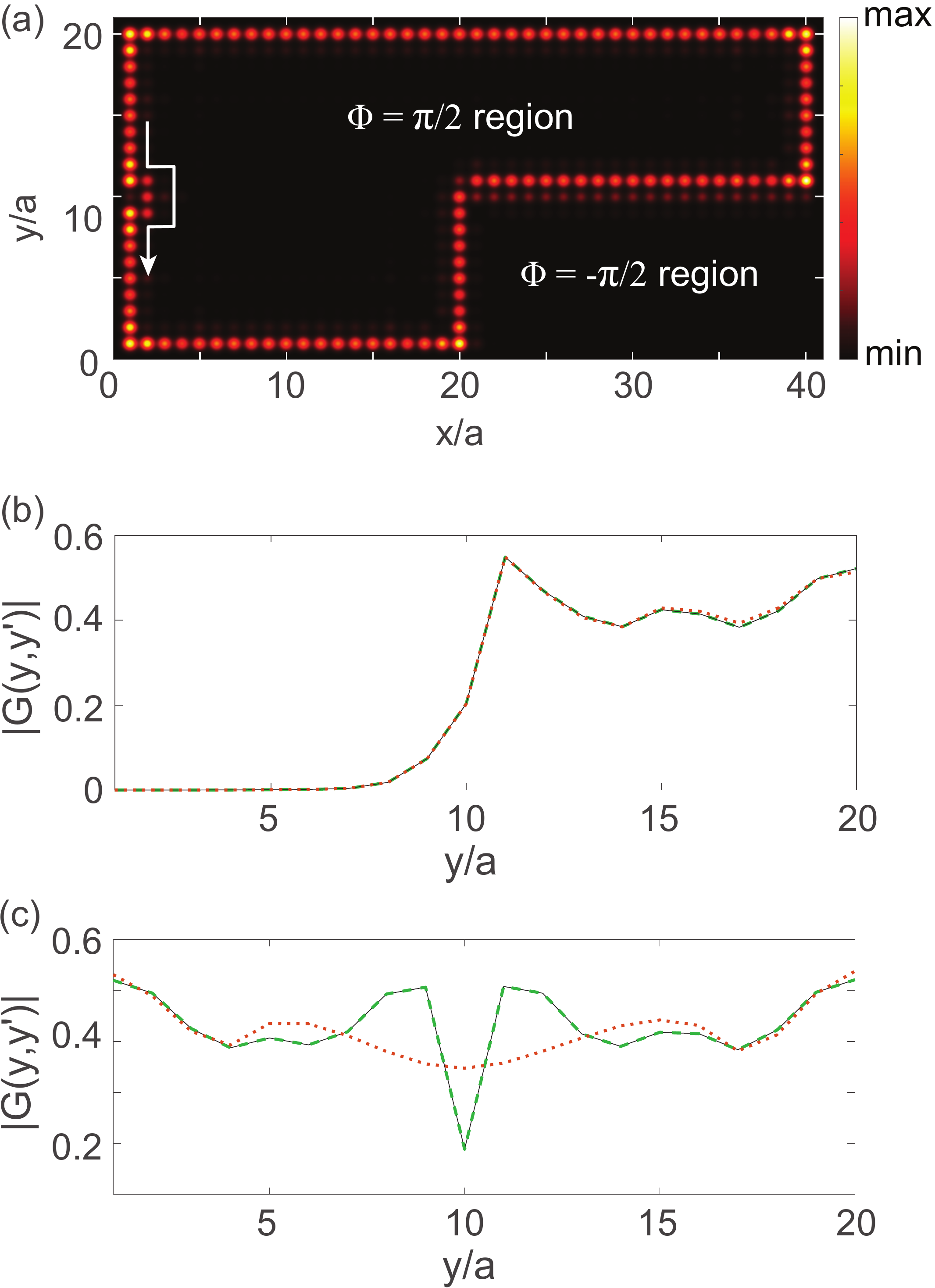}
	\caption{Analogy between the Green's function and a chiral edge state circumventing a defect in a topological square lattice. (a) The Green's function calculated with $z=1.85g$ and the source at $x=a,y=10a$. Solid, dashed, and dotted lines show the results of our auxiliary eigenvalue approach, the recursive Green's function, and the bilinear expansion, respectively. (b,c) The values of the Green's function along $y$ at $x=40a$ and $a$, respectively.}\label{fig:topo}
\end{figure}

Finally, we reveal the most intriguing connection between the Green's function and a defect state, i.e., in a 2D topological lattice with a chiral edge state. This system breaks Lorentz reciprocity \cite{Landau,ge_reciprocity2016}, which can be achieved in an electronic system by imposing a magnetic field. An analogy can be introduced to photonic systems by imposing an artificial gauge field, achieved experimentally by asymmetric couplings between two neighboring lattice sites on a tight-binding lattice \cite{hafezi_robust_2011,hafezi_imaging_2013,bandres_topological_2018,zhao_non-hermitian_2019}: while the couplings in both the $x$ and $y$ directions still have the same amplitude, their phases are now different. Here we consider a square lattice (Figure~\ref{fig:topo}) with uniform vertical coupling and horizontal asymmetric couplings of the same amplitude $g$, realizing a Landau gauge with a $\pi/2$-flux through the smallest plaquette \cite{hafezi_robust_2011,hafezi_imaging_2013}. Its bottom right corner is pierced by the opposite flux, and an on-site potential shift of $-2g$ is also introduced to decouple it from the rest of the system.

Due to its sublattice symmetry \cite{Hasan}, the energy spectrum of the main region is symmetric about $z=0$, set at the value of the identical on-site potential. It has an edge band with CCW chiral edge states in $z/g\in[1.08, 2.61]$, with their CW counterparts in $z/g\in[-2.61, -1.08]$. We probe the response of the system by placing a point source at $x=a,y=10a$, where $a=1$ is the lattice constant. As Figure~\ref{fig:topo}(a) shows, the Green's function with $z$ at the middle of the CCW chiral edge band displays the same characteristic as the eigenstates of $\mathcal{L}$, i.e., localized near the edge of the topological region. Our auxiliary eigenvalue approach gives identical result as the recursive Green's function method [solid and dashed lines in Figs.~\ref{fig:topo}(b) and (c)], and the Green's function displays two noticeable features when compared with the bilinear expansion: while only a small number ($\sim$20) of edge states are needed to capture the Green's function on the opposite (right) edge of the system (dotted line), they cannot describe the behavior of the Green's function near the source, where it displays a local minimum.

One way to understand this behavior is offered by our auxiliary eigenvalue approach, where the Green's function is effectively a defect state perturbed by the $\delta$-function potential [see Equation~(\ref{eq:defect})]. In this point of view, the local minimum of the Green's function is the consequence of the topological protection of the chiral edge states: it circumvents this defect as can be clearly seen in Figure~\ref{fig:topo}(a). This change, of course, creates a local perturbation so strong that can only be captured by a large number of the unperturbed eigenstates of $\mathcal{L}$ in the bilinear expansion, including even the chiral edge states in the opposite direction that are much different in energy (not shown). The far-field, on the other hand, is not affected by this local perturbation, as we have seen in Figure~\ref{fig:topo}(b).  We note that the Green's function does not display a local minimum at all energies: if $z$ is almost on resonance with an unperturbed eigenstate in the edge band, the Green's function is given essentially by this single chiral edge mode, which goes through the defect created by the $\delta$-function potential with little scattering.

\section{Discussion and conclusions}

In summary, we have introduced in this work a new perspective of the time-independent Green's function, which is obtained as a single eigenstate of an auxiliary eigenvalue formulation that embodies a defect state created by the $\delta$-function potential. The height of the $\delta$-function potential is determined by the inverse value of the Green's function at the source position, which is given directly in the form of a generalized eigenvalue problem given by Equation~(\ref{eq:G_new_def}). It is the only well behaved and finite eigenvalue, easy to identify numerically. Therefore, our approach differs both conceptually and computationally from previous investigations of eigenstates of a $\delta$-function potential, which were not related to the Green's function.

The uniqueness of the eigenstate that gives the Green's function in our approach should be distinguished from the single bound state in an attractive 1D $\delta$-function potential \cite{Gottfried}. In our case the $\delta$-function potential can be repulsive or attractive in a Hermitian system [see Figure~\ref{fig:Schrodinger}(a)], and it becomes complex in general in non-Hermitian systems. Furthermore, the eigenstate in our approach does not depend on the original potential included in the operator $\cal L$, while the aforementioned bound state assumes a vacuum background. Finally, our eigenstate exists in higher dimensions as well, with clearly different boundary conditions and spatial profiles from a bound state [see Figure~\ref{fig:topo}(a), for example].

We have also verified the Green's function obtained in our method by comparing both to analytical results when available and to two other frequently used numerical methods, i.e., the recursive Green's function method and the bilinear expansion in the basis of the system's eigenstates. At an EP, where a perturbative treatment of the bilinear expansion becomes necessary, our method is still robust as seen from the examples in two $\mathcal{PT}$-systems. Our method also gives identical result to the recursive Green's function method, implemented by finite difference on the same tight-binding lattice.

Our defect-state approach can also be applied to more numerically demanding cases, e.g., in the study of diffusive transport and wave localization \cite{Beenakker1997,Imry2008}. We have investigated disordered quasi-1D waveguides over 100 wavelengths long and with over 60 transverse channels (not shown). In these cases, the memory storage for the recursive Green's function may become an issue, when the values of the Green's function in the interior of the waveguides are also computed for light-matter interaction or laser emissions. This is because the Green's matrix obtained from the recursive procedure mingles the values of the Green's function generated by sources placed across the entire system. The defect-state approach excels in this regard, because each source is treated independently. We should also mention that this advantage becomes less noticeable or even a disadvantage if the Green's function needs to be evaluated with the source at many different locations.

More importantly, our approach offers a previously unexplored physical insight that both the recursive Green's function method and the bilinear expansion lack, i.e., the linkage between the Green's function and a defect state. We have exemplified an intriguing manifestation of this linkage using a topological chiral edge state, where the local minimum of the Green's function is analogous to a chiral edge state circumventing a boundary defect.  Therefore, even though our discussions have focused on the Helmholtz equation for scalar optical waves, we expect this new perspective and the physical insight it offers to have important applications in other related fields as well, including condensed matter systems, acoustics, electronic circuits and so on.

\medskip
\textbf{Acknowledgements} \par 
We thank Douglas Stone, Yidong Chong and Azriel Genack for helpful discussions. This project is supported by the NSF under Grant No. PHY-1847240.
\medskip

\appendix

\section{Reciprocity in the auxiliary eigenvalue approach}
\label{sec:reciproc}

We derive some relations between the Green's functions $G(\bm{r},\bm{r'})$ and $G(\bm{r'},\bm{r})$, based on the symmetries of $\mathcal{L}$. We start from a matrix description in which the Green's function operator $G(z)$ is defined from $G(\bm{r},\bm{r'};z) = \langle \bm{r}|G(z)|\bm{r'}\rangle$. We use the same symbol $\mathcal{L}$ for the matrix representation of the system. The Green's function satisfies the following equation:
\begin{eqnarray}
	[z\bm{1} - \mathcal{L}]G(z) = \bm{1}, \hspace{0.5cm} G(z)[z\bm{1} - \mathcal{L}] = \bm{1},
	\label{eq:left_right_inverse}
\end{eqnarray}
where $\bm{1}$ is the identity matrix. Here, we identify the Green's function as being both the right and left matrix inverses of $[z\bm{1} - \mathcal{L}]$. Considering that in non-Hermitian physics, there is a distinction between left and right sets of eigenvectors \cite{Noether_nonHermitian}, one may attempt to distinguish between the left and right Green's functions. However, it can be shown that they are identical, since $[z\bm{1} - \mathcal{L}]$ is a matrix of full rank. It is possible to obtain relations between $G(z)$ and $[G(z)]^T$, or $[G(z)]^\dagger$, when performing matrix transposition or hermitian conjugation on Equation (\ref{eq:left_right_inverse}).

We start by analyzing the case $\mathcal{L}^T = \mathcal{L}$: matrix transposition of Equation (\ref{eq:left_right_inverse}) leads to the relation $G(z)^T = G(z)$, or more explicitly $G(\bm{r'},\bm{r};z) = G(\bm{r},\bm{r'};z)$, which is the usual Lorentz reciprocity condition. This relation holds when $\mathcal{L}$ is a symmetric matrix, describing a system with symmetric couplings and hence without an effective magnetic field. It is valid for non-Hermitian systems with this property as well, such as the case with a non-Hermitian flatband discussed in Figure 7 of the main text.

In a general Hermitian system, $\mathcal{L}^\dagger = \mathcal{L}$, and uniqueness of the matrix inverse demands that $[G(z)]^T = [G(z^*)]^*$, or $G(\bm{r'},\bm{r};z) = [G(\bm{r},\bm{r'};z^*)]^*$. It is customary to state this relation in terms of Green's functions $G^{(\pm)}$ with \textit{outgoing} and \textit{incoming} boundary conditions, respectively. This nomenclature is specially useful if $z = k + i\eta$, with $\eta>0$ small compared to $k$:

\be
G^{(\pm)}(\bm{r},\bm{r'};k) \equiv \lim\limits_{\eta \to 0^+} G(\bm{r},\bm{r'};k \pm i\eta).
\ee
Therefore, we can write the reciprocity condition as $G^{(\pm)}(\bm{r'},\bm{r};k) = [G^{(\mp)}(\bm{r},\bm{r'};k)]^*$. The solutions $G^{(\pm)}$ are oftentimes referred to as the \textit{retarded} and \textit{advanced} Green's functions, inspired from the association to sources generating outgoing or incoming waves. Systems that are non-Hermitian and do not have symmetric couplings lead to more complicated reciprocity relations, involving the Green's functions of two different systems. This is expected, since the scattering properties derived from the Green's functions of systems related by a time-reversal transformation are nontrivial and yet to be understood. Reference \cite{Jose_scaling2019} provides interesting scattering relations between two ``time-reversal partners'' in 1D and quasi 1D systems.

Our Green's function approach for time-reversal or non-Hermitian symmetric systems lead to a relation between the single eigenstates:

\be
\frac{\psi_0(\bm{r},\bm{r'})}{\lambda_0(\bm{r'}) \psi_0(\bm{r'},\bm{r'})} = \frac{\psi_0(\bm{r'},\bm{r})}{\lambda_0(\bm{r}) \psi_0(\bm{r},\bm{r})},
\ee
or
\be
\psi_0(\bm{r'},\bm{r}) = \psi_0(\bm{r},\bm{r'})\left[\frac{\lambda_0(\bm{r})\psi_0(\bm{r},\bm{r})}{\lambda_0(\bm{r'})\psi_0(\bm{r'},\bm{r'})}\right],
\ee
whereas for the general non-Hermitian problem, we find
\be
\bar{\psi}_0(\bm{r'},\bm{r}) = [\psi_0(\bm{r},\bm{r'})]^*\left[\frac{\bar{\lambda}_0(\bm{r})\bar{\psi}_0(\bm{r},\bm{r})}{\lambda_0(\bm{r'})\psi_0(\bm{r'},\bm{r'})}\right],
\ee
where we used the notation $\bar{\psi}_0(\bm{r},\bm{r'},z) \equiv \psi_0(\bm{r},\bm{r'},z^*)$. The same result is obtained if we identify $\bar{\psi}_0$ with the advanced solution and $\psi_0$ with the retarded solution, or vice-versa.

\section{Perturbation theory at an EP}
\label{sec:pert}
Below we review a generic procedure to approach effective $\mathcal{PT}$-symmetric systems with the outgoing boundary condition based on the coupled-mode theory. In particular, we describe a photonic molecule that consists on identical components $\mathbb{L}$ and $\mathbb{R}$ that support localized states $\psi_l(x)$ and $\psi_r(x)$ in each component. The system can be, for instance, two identical whispering gallery resonators that are brought to proximity, such that their modes overlap and produce a pair of bonding and antibonding states. Alternatively, one can think of a pair of half-wavelength cavities coupled by a DBR, which is the system discussed in Section 2.2 of the main text. Note that due to the outgoing boundary condition, such systems are not truly \pt-symmetric, because time reversal changes the outgoing boundary condition to the incoming boundary condition. However, if the states are highly localized as in our discussion below, the minute flux in or out of the system can be omitted in the discussion of the coupled mode theory.

The left component satisfies the Helmholtz equation:
\begin{align}
	\left[ \partial_x^2 + \epsilon_l(x) k_0^2 \right] \psi_l(x) &= 0,\quad (x\in\mathbb{L}) \label{eq:L1} \\
	\left[ \partial_x^2 + k^2 \right] \psi_l(x) &= 0. \quad (x\in \text{elsewhere})\label{eq:L2}
\end{align}
Here $k$ is the real-valued external frequency at which the Green's function is going to be evaluated, and $k_0\sim k$ is one complex-valued CF frequency. The right component is defined similarly, and the localized states are normalized by $\int_{x\in\mathbb{L},\mathbb{R}} \epsilon_{l,r} \psi_{l,r}^2 dx =1$, respectively. The composite system is defined by $\epsilon_0(x)=\epsilon_{l}(x)+\epsilon_r(x)$ and satisfies
\begin{align}
	\left[ \partial_x^2 + \epsilon_0(x)\tilde{k}^2 \right] \tilde{\psi}(x) &= 0,\quad (x\in\mathbb{L}\cup\mathbb{R})\label{eq:W1}\\
	\left[ \partial_x^2 + k^2 \right] \tilde{\psi}(x) &= 0, \quad (x\in \text{elsewhere})\label{eq:W2}
\end{align}
where $\tilde{k}$ is the CF frequency of a ``supermode" 	
\be
\tilde{\psi}(x) = a_l \psi_l(x) + a_r\psi_r(x)	\label{eq:sum}
\ee
to be determined. We note that $\tilde{\psi}(x)$ automatically satisfies the outgoing boundary condition at the two ends of the composite system due to the same property of both $\psi_{l,r}(x)$.

We make the following assumption about the system: the spatial overlapping of the localized states in one cavity is very weak, i.e., $\int_{x\in\mathbb{L},\mathbb{R}} \epsilon_{l,r} \psi_l\psi_r dx\equiv g\;(|g|\ll1)$. In addition, we note that the frequency shift due to $|\int_{x\in\mathbb{R}} \epsilon_r \psi_l^2 dx| = |\int_{x\in\mathbb{L}} \epsilon_l \psi_r^2 dx|$ is even smaller, which is then neglected in our analysis below.

To derive the coupled-mode theory, we first insert Equation~(\ref{eq:sum}) into Eqs. (\ref{eq:W1}), (\ref{eq:W2}) and simplify the result using Eqs. (\ref{eq:L1}), (\ref{eq:L2}) for $\psi_l(x)$ and their counterparts for $\psi_r(x)$. Then, we multiply the resulting equation by either $\psi_l(x)$ or $\psi_r(x)$ and perform an integration in the left and right component of the system respectively, obtaining in this way two independent equations:

\be
\begin{bmatrix}
	\tilde{k}^2-k_0^2 & g\tilde{k}^2 - tk^2\\
	g\tilde{k}^2-tk^2  & \tilde{k}^2-k_0^2
\end{bmatrix}
\begin{bmatrix}
	a_l \\
	a_r
\end{bmatrix}=0.
\ee
It is a system of equations for the unknown amplitudes $a_{l,r}$ and frequency $\tilde{k}$, where $t$ is defined by $\int_{x\in\mathbb{L}}\psi_r^2dx=\int_{x\in\mathbb{R}}\psi_l^2dx$. Since the matrix operator is symmetric and has identical diagonal elements, a zero-determinant condition implies that $a_r = \pm a_l$, giving rise to the \textit{symmetric} and \textit{anti-symmetric} modes $\tilde{\psi}_S(x)$ and $\tilde{\psi}_A(x)$ respectively, with frequencies:
\begin{eqnarray}
	\tilde{k}_{S,A}^2 = \frac{k_0^2\pm tk^2}{1 \pm g}.
\end{eqnarray}
With the aforementioned approximations $|g| \ll 1$ and $k_0\sim k$, these frequencies reduce to
\begin{eqnarray}
	\tilde{k}_S^2 \approx (1+t-g) k_0^2, \hspace{.5cm} \tilde{k}_A^2 \approx \left(1-t+g\right) k_0^2,\label{ap6}
\end{eqnarray}
and their splitting equal to $\tilde{k}_S^2 - \tilde{k}_A^2 = 2(t-g) k_0^2\equiv2\Delta k_0^2$.

Now, we introduce a $\mathcal{PT}$-symmetric perturbation $i\epsilon_1(x)$, which is an odd function of $x$. The effective \pt-symmetric system satisfies
\begin{equation}
	[\partial_x^2 + (\epsilon_0(x) + i \epsilon_1(x))q^2]\psi(x) = 0, \label{ap8}
\end{equation}
where $q$ is the perturbed CF frequency to be determined together with the amplitudes $A$ and $S$ in the expansion of the perturbed eigenstate $\psi(x) = A\tilde{\psi}_A(x) + S\tilde{\psi}_S(x)$. Following an analogous procedure to the one described above, we obtain again a system of equations that can be represented by the following matrix equation for $A$, $S$:
\begin{equation}
	\begin{bmatrix}
		q^2 - \tilde{k}_S^2 & i Cq^2  \\
		i Cq^2 & q^2 - \tilde{k}_A^2
	\end{bmatrix}\begin{bmatrix}
		S\\
		A
	\end{bmatrix} = 0,
	\label{ap9}
\end{equation}
where $C \equiv \int \epsilon_1(x)\tilde{\psi}_A(x)\tilde{\psi}_S(x) dx$ is integrated over the whole system. We note that this equation and the following perturbative results apply when $\epsilon_0(x)$ is complex, i.e., the system does not need to have balanced gain and loss to be effectively $\cal{PT}$ symmetric \cite{Guo}.

To rewrite it in a more familiar form, we note that since the \pt-symmetric perturbation is assumed to be weak, we approximate $Cq^2$ by $Ck_0^2$ that leads to
\begin{equation}
	\tilde{H}
	\begin{bmatrix}
		S\\
		A
	\end{bmatrix}
	\equiv
	\begin{bmatrix}
		k_S^2 & -iCk_0^2  \\
		-iCk_0^2  & k_A^2
	\end{bmatrix}
	\begin{bmatrix}
		S\\
		A
	\end{bmatrix}
	= q^2
	\begin{bmatrix}
		S\\
		A
	\end{bmatrix}
	.\label{ap09}
\end{equation}
The perturbed frequency $q$ is then found to be:
\begin{equation}
	q^2_\pm \approx k_0^2 \left(1\pm \sqrt{\Delta^2 - C^2 }\right),\label{ap10}
\end{equation}
where we have used the forms of $\tilde{k}_{S,A}^2$ from Equation~(\ref{ap6}). The EP then emerges when the $C^2 = \Delta^2$, where $q_\pm \equiv k_\text{EP}  \approx k_0$. In the main text we have introduced $\varsigma\equiv\sqrt{1-\beta^2}$ and $\beta\equiv C/\Delta$.

This system of equations can be transformed to the basis of left- and right-components, in terms of the amplitudes $a_l$, $a_r$:

\begin{equation}
	H
	\begin{bmatrix}
		a_l\\
		a_r
	\end{bmatrix}
	\equiv
	k_0^2\begin{bmatrix}
		1- iC & \Delta \\
		\Delta  & 1+iC
	\end{bmatrix}
	\begin{bmatrix}
		a_l\\
		a_r
	\end{bmatrix}
	=
	q^2\begin{bmatrix}
		a_l\\
		a_r
	\end{bmatrix}.\label{eq:CMT_appendix}
\end{equation}
It has the most familiar form of a $\mathcal{PT}$-symmetric Hamiltonian. This is the model that we have used to describe the photonic molecule in Section 2.2.

The perturbative result of the Green's function given by Equation~(16) in the main text is derived from the following expression:
\begin{widetext}
\begin{align}
	G(x,x';k) = \frac{\epsilon_0(x')}{(k^2-k_0^2)^2-\Delta^2\varsigma^2}\{(k^2&-k_0^2)[\psi_l(x')\psi_l(x)+\psi_r(x')\psi_r(x)]\, \nonumber\\
	&-\Delta k_0^2[\,i\beta[\psi_l(x')\psi_l(x)-\psi_r(x')\psi_r(x)]-[\psi_l(x')\psi_r(x)+\psi_r(x')\psi_l(x)]\,]\}\label{eq:G_full}
\end{align}
\end{widetext}
by dropping terms proportional to $\varsigma^2$, including that in $\beta=\sqrt{1-\varsigma^2}\approx1$. We note that Equation~(\ref{eq:G_full}) includes terms to all orders of $\varsigma$. Here it is unnecessary to invoke the Jordan vector $J(x)$ that completes the Hilbert space at the EP \cite{Chen2020}, but to compare with the result given in Reference~\cite{Pick2017}, we note that it can be chosen as
\be
J(x) = \frac{i\psi_l(x)+\psi_r(x)}{2\sqrt{\Delta}k_0}e^{-i\pi/4},
\ee
which satisfies
\be
[H-k_\text{EP}^2\mathbf{1}]J(x) = \psi_\text{EP}(x)
\ee
with the coalesced eigenstate
\be
\psi_\text{EP}(x) = 2\sqrt{\Delta}k_0e^{-i\pi/4}[\psi_l(x)+i\psi_r(x)]
\ee
at the EP. Here $\mathbf{1}$ is the identity matrix, and $\psi_\text{EP}(x)$ is normalized differently from the main text to satisfy $(\psi_\text{EP},J)=1$ [see the definition of the inner product given by Equation~(14) in the main text]. We further note that the \pt-symmetric Hamiltonian given by Equation~(\ref{eq:CMT_appendix}) above is symmetric, and hence it is unnecessary to distinguish left and right eigenstates (and Jordan vectors) because they are identical. Using $J(x)$ and $\psi_\text{EP}(x)$ specified above, we find that the Green's function derived in Reference~\cite{Pick2017} at the EP agrees with our result given by Equation~(16) in the main text, once the additional factor $\epsilon_0(x')$ is accounted for that originates from rewriting the Helmholtz equation as Equation~(6) in the main text.

\section{The recursive Green's function}
\label{sec:RGF}

Below we briefly review the recursive Green's function method and our implementation of it using finite difference. Assuming that the Green's function defined by Equation~(1) has been obtained at every grid point for a quasi-1D waveguide with $N-1$ columns and $N_y$ rows, we can write down a Green's matrix $\bm{G}^{(N-1)}$ that satisfies
\be
\left[z\bm{1}^{(N-1)}-\bm{L}^{(N-1)}\right]\bm{G}^{(N-1)}=\bm{1}^{(N-1)}.\label{eq:RGF_N-1}
\ee
Its individual element $\bm{G}^{(N-1)}_{\underline{i,j}}$ gives the value of the Green's function at point $i$ when the source is placed at point $j$. We reserve the notation $\bm{G}^{(N-1)}_{i,j}$ without the underscore below the subscripts for the $N_y\times N_y$ block of the Green's matrix evaluated in the $i$th column with the source in the $j$th column, and we apply the same notation to other matrices as well. The $(N-1)N_y$ grid points are labeled from the first to the last point in the first column, continued in the same way onto the second and remaining columns. $\bm{1}^{(N-1)}$ is the identity matrix with $(N-1)N_y$ rows and columns, and $\bm{L}^{(N-1)}$ is the finite difference implementation of the operator $\mathcal{L}$ in Equation~(1), again with $(N-1)N_y$ rows and columns.

To calculate the Green's matrix with an additional column added from the right, we separate $\bm{L}^{(N)}$ into $\bm{L}^0 + \bm{V}$ and construct an ancillary matrix $\bm{G}^0$, where
\be
\bm{L}^0 \equiv
\begin{pmatrix}
	\bm{L}^{(N-1)} & \\
	& \bm{L}^{(N)}_{N,N}
\end{pmatrix}, \quad
\bm{G}^0 \equiv
\begin{pmatrix}
	\bm{G}^{(N-1)} & \\
	& \bm{G}^{0}_{N,N}
\end{pmatrix}.\nonumber
\ee
$\bm{V}$ only contains the couplings between the $(N-1)$th and $N$th columns in $\bm{L}^{(N)}$, and
\be
\left[z\bm{1}^{(1)}-\bm{L}^{(N)}_{N,N}\right]\bm{G}^{0}_{N,N}=\bm{1}^{(1)}
\ee
defines the Green's matrix $\bm{G}^{0}_{N,N}$ of the isolated $N$th column. It is straightforward to show that
\be
\left[z\bm{1}^{(N)}-\bm{L}^{0}\right]\bm{G}^{0}=\bm{1}^{(N)}
\ee
and we derive the well known Dyson's equation
\be
\bm{G}^{(N)} = \bm{G}^0 + \bm{G}^{(N)}\bm{V}\bm{G}^0,\label{eq:RGF0}
\ee
where $\bm{G}^{(N)}$ is defined by
\be
\left[z\bm{1}^{(N)}-\bm{L}^{(N)}\right]\bm{G}^{(N)}=\bm{1}^{(N)}
\ee
similar to Equation~(\ref{eq:RGF_N-1}). Equation~(\ref{eq:RGF0}) can be verified directly by multiplying $[z\bm{1}^{(N)}-\bm{L}^{(N)}]$ from the left.

Below we drop the superscript of $\bm{G}^{(N)}$. Equation~(\ref{eq:RGF0}) differs from its more familiar form in mesoscopic physics \cite{Sols_theory_1989}, i.e.,
\be
\bm{\tilde{G}} = \bm{\tilde{G}}^0 + \bm{\tilde{G}}^{0}\bm{V}\bm{\tilde{G}},
\ee
because there the Green's matrix is defined by $\bm{\tilde{G}}[z\bm{1}^{(N)}-\bm{L}^{(N)}]=\bm{1}^{(N)}$ instead. We have already proven the equivalence between the left and right matrix inverses of $[z\bm{1} - \mathcal{L}]$ in Appendix \ref{sec:reciproc}, thus $\tilde{\bm{G}}$ and $\bm{G}$ are identical.


Using Equation~(\ref{eq:RGF0}), the recursive relations between the blocks of the Green's matrices $\bm{G}_{i,j}$ connecting the two boundary layers (i.e., the 1st and $N$th columns) can be derived:
\begin{eqnarray}
	&\bm{G}_{NN} = \;A^{-1}, \\
	&\bm{G}_{1N}\; =\; \bm{G}^0_{1,N-1}\bm{V}_{N-1,N}A^{-1}, \\
	&\bm{G}_{N1}\; =\; \bm{G}^0_{N,N}\bm{V}_{N,N-1}\bm{G}^0_{N-1,1}, \\
	&\bm{G}_{11}\;\; =\; \bm{G}^0_{1,1} + \bm{G}_{1,N}\bm{V}_{N,N-1}\bm{G}^0_{N-1,1}.
\end{eqnarray}

\noindent A single matrix inversion for
\be
A = z\bm{1}^{(1)}-\bm{L}^{(N)}_{N,N} - V_{N,N-1}\bm{G}^0_{N-1,N-1}V_{N-1,N}
\ee
is needed when adding one layer from the right, and its last term is often referred to as the self-energy \cite{thouless_conductivity_1981}. Note that besides the operator $L^{(N)}$, all the pieces of information needed for this recursive scheme are the same four matrix for the system with $(N-1)$ layers. As a result, the recursive Green's function method is highly efficient to treat transport problems, which does not require the knowledge of the Green's function inside the system. If we need the full Green's matrix to investigate, for example, light-matter interaction in the bulk of the system, more steps are required in the recursive Green's function, which scale as $N^2$ and are more intense numerically.


\begin{thebibliography}{53}%
	\makeatletter
	\providecommand \@ifxundefined [1]{%
		\@ifx{#1\undefined}
	}%
	\providecommand \@ifnum [1]{%
		\ifnum #1\expandafter \@firstoftwo
		\else \expandafter \@secondoftwo
		\fi
	}%
	\providecommand \@ifx [1]{%
		\ifx #1\expandafter \@firstoftwo
		\else \expandafter \@secondoftwo
		\fi
	}%
	\providecommand \natexlab [1]{#1}%
	\providecommand \enquote  [1]{``#1''}%
	\providecommand \bibnamefont  [1]{#1}%
	\providecommand \bibfnamefont [1]{#1}%
	\providecommand \citenamefont [1]{#1}%
	\providecommand \href@noop [0]{\@secondoftwo}%
	\providecommand \href [0]{\begingroup \@sanitize@url \@href}%
	\providecommand \@href[1]{\@@startlink{#1}\@@href}%
	\providecommand \@@href[1]{\endgroup#1\@@endlink}%
	\providecommand \@sanitize@url [0]{\catcode `\\12\catcode `\$12\catcode
		`\&12\catcode `\#12\catcode `\^12\catcode `\_12\catcode `\%12\relax}%
	\providecommand \@@startlink[1]{}%
	\providecommand \@@endlink[0]{}%
	\providecommand \url  [0]{\begingroup\@sanitize@url \@url }%
	\providecommand \@url [1]{\endgroup\@href {#1}{\urlprefix }}%
	\providecommand \urlprefix  [0]{URL }%
	\providecommand \Eprint [0]{\href }%
	\providecommand \doibase [0]{https://doi.org/}%
	\providecommand \selectlanguage [0]{\@gobble}%
	\providecommand \bibinfo  [0]{\@secondoftwo}%
	\providecommand \bibfield  [0]{\@secondoftwo}%
	\providecommand \translation [1]{[#1]}%
	\providecommand \BibitemOpen [0]{}%
	\providecommand \bibitemStop [0]{}%
	\providecommand \bibitemNoStop [0]{.\EOS\space}%
	\providecommand \EOS [0]{\spacefactor3000\relax}%
	\providecommand \BibitemShut  [1]{\csname bibitem#1\endcsname}%
	\let\auto@bib@innerbib\@empty
	\bibitem [{\citenamefont {Green}(1871)}]{green_book}%
	\BibitemOpen
	\bibfield  {author} {\bibinfo {author} {\bibfnamefont {G.}~\bibnamefont
			{Green}},\ }\href@noop {} {\emph {\bibinfo {title} {{Mathematical} papers of
				the late {G}eorge {G}reen}}}\ (\bibinfo  {publisher} {Macmillan},\ \bibinfo
	{address} {London},\ \bibinfo {year} {1871})\BibitemShut {NoStop}%
	\bibitem [{\citenamefont {Stakgold}\ and\ \citenamefont
		{Holst}(2011)}]{Stakgold2011}%
	\BibitemOpen
	\bibfield  {author} {\bibinfo {author} {\bibfnamefont {I.}~\bibnamefont
			{Stakgold}}\ and\ \bibinfo {author} {\bibfnamefont {M.}~\bibnamefont
			{Holst}},\ }\href {http://repositorio.unan.edu.ni/2986/1/5624.pdf} {\emph
		{\bibinfo {title} {{Green's functions and boundary value problems}}}},\
	\bibinfo {edition} {3rd}\ ed.,\ Pure and Applied Mathematics\ (\bibinfo
	{publisher} {John Wiley {\&} Sons},\ \bibinfo {address} {Hoboken},\ \bibinfo
	{year} {2011})\BibitemShut {NoStop}%
	\bibitem [{\citenamefont {Melnikov}\ and\ \citenamefont
		{Borodin}(2017)}]{Melnikov2017}%
	\BibitemOpen
	\bibfield  {author} {\bibinfo {author} {\bibfnamefont {Y.~A.}\ \bibnamefont
			{Melnikov}}\ and\ \bibinfo {author} {\bibfnamefont {V.~N.}\ \bibnamefont
			{Borodin}},\ }\href@noop {} {\emph {\bibinfo {title} {{Green's
					Functions}}}},\ \bibinfo {series} {Developments in Mathematics}\ No.~\bibinfo
	{number} {48}\ (\bibinfo  {publisher} {Springer},\ \bibinfo {address}
	{Cham},\ \bibinfo {year} {2017})\BibitemShut {NoStop}%
	\bibitem [{\citenamefont {Morse}\ and\ \citenamefont
		{Feshbach}(1953)}]{Morse1953}%
	\BibitemOpen
	\bibfield  {author} {\bibinfo {author} {\bibfnamefont {P.~M.}\ \bibnamefont
			{Morse}}\ and\ \bibinfo {author} {\bibfnamefont {H.}~\bibnamefont
			{Feshbach}},\ }\href@noop {} {\emph {\bibinfo {title} {Methods of Theoretical
				Physics}}},\ Vol.~\bibinfo {volume} {1}\ (\bibinfo  {publisher}
	{McGraw-Hill},\ \bibinfo {address} {New York},\ \bibinfo {year}
	{1953})\BibitemShut {NoStop}%
	\bibitem [{\citenamefont {Schwinger}(1998)}]{schwinger_book}%
	\BibitemOpen
	\bibfield  {author} {\bibinfo {author} {\bibfnamefont {J.}~\bibnamefont
			{Schwinger}},\ }\href@noop {} {\emph {\bibinfo {title} {Particles, {Sources},
				and {Fields}}}}\ (\bibinfo  {publisher} {Perseus},\ \bibinfo {address}
	{Reading, Mass},\ \bibinfo {year} {1998})\BibitemShut {NoStop}%
	\bibitem [{\citenamefont {Rickayzen}(1980)}]{Rickayzen1980}%
	\BibitemOpen
	\bibfield  {author} {\bibinfo {author} {\bibfnamefont {G.}~\bibnamefont
			{Rickayzen}},\ }\href@noop {} {\emph {\bibinfo {title} {Green's functions and
				Condensed Matter}}}\ (\bibinfo  {publisher} {Academic Press},\ \bibinfo
	{address} {New York},\ \bibinfo {year} {1980})\BibitemShut {NoStop}%
	\bibitem [{\citenamefont {Economou}(2007)}]{Economou2007}%
	\BibitemOpen
	\bibfield  {author} {\bibinfo {author} {\bibfnamefont {E.}~\bibnamefont
			{Economou}},\ }\href {https://doi.org/10.1002/pc.23732} {\emph {\bibinfo
			{title} {{Green's functions in Quantum Physics}}}},\ \bibinfo {edition}
	{3rd}\ ed.\ (\bibinfo  {publisher} {Springer},\ \bibinfo {year}
	{2007})\BibitemShut {NoStop}%
	\bibitem [{\citenamefont {Davy}\ \emph {et~al.}(2015)\citenamefont {Davy},
		\citenamefont {Shi}, \citenamefont {Wang}, \citenamefont {Cheng},\ and\
		\citenamefont {Genack}}]{Davy2015}%
	\BibitemOpen
	\bibfield  {author} {\bibinfo {author} {\bibfnamefont {M.}~\bibnamefont
			{Davy}}, \bibinfo {author} {\bibfnamefont {Z.}~\bibnamefont {Shi}}, \bibinfo
		{author} {\bibfnamefont {J.}~\bibnamefont {Wang}}, \bibinfo {author}
		{\bibfnamefont {X.}~\bibnamefont {Cheng}},\ and\ \bibinfo {author}
		{\bibfnamefont {A.~Z.}\ \bibnamefont {Genack}},\ }\bibfield  {title}
	{\bibinfo {title} {Transmission {Eigenchannels} and the {Densities} of
			{States} of {Random} {Media}},\ }\href
	{https://doi.org/10.1103/PhysRevLett.114.033901} {\bibfield  {journal}
		{\bibinfo  {journal} {Phys. Rev. Lett.}\ }\textbf {\bibinfo {volume} {114}},\
		\bibinfo {pages} {033901} (\bibinfo {year} {2015})}\BibitemShut {NoStop}%
	\bibitem [{\citenamefont {Lin}\ \emph {et~al.}(2016)\citenamefont {Lin},
		\citenamefont {Pick}, \citenamefont {Lon{\v{c}}ar},\ and\ \citenamefont
		{Rodriguez}}]{Lin2016}%
	\BibitemOpen
	\bibfield  {author} {\bibinfo {author} {\bibfnamefont {Z.}~\bibnamefont
			{Lin}}, \bibinfo {author} {\bibfnamefont {A.}~\bibnamefont {Pick}}, \bibinfo
		{author} {\bibfnamefont {M.}~\bibnamefont {Lon{\v{c}}ar}},\ and\ \bibinfo
		{author} {\bibfnamefont {A.~W.}\ \bibnamefont {Rodriguez}},\ }\bibfield
	{title} {\bibinfo {title} {{Enhanced Spontaneous Emission at Third-Order
				Dirac Exceptional Points in Inverse-Designed Photonic Crystals}},\ }\href
	{https://doi.org/10.1103/PhysRevLett.117.107402} {\bibfield  {journal}
		{\bibinfo  {journal} {Phys. Rev. Lett.}\ }\textbf {\bibinfo {volume} {117}},\
		\bibinfo {pages} {107402} (\bibinfo {year} {2016})}\BibitemShut {NoStop}%
	\bibitem [{\citenamefont {Pick}\ \emph {et~al.}(2017)\citenamefont {Pick},
		\citenamefont {Zhen}, \citenamefont {Miller}, \citenamefont {Hsu},
		\citenamefont {Hernandez}, \citenamefont {Rodriguez}, \citenamefont
		{Soljačić},\ and\ \citenamefont {Johnson}}]{Pick2017}%
	\BibitemOpen
	\bibfield  {author} {\bibinfo {author} {\bibfnamefont {A.}~\bibnamefont
			{Pick}}, \bibinfo {author} {\bibfnamefont {B.}~\bibnamefont {Zhen}}, \bibinfo
		{author} {\bibfnamefont {O.~D.}\ \bibnamefont {Miller}}, \bibinfo {author}
		{\bibfnamefont {C.~W.}\ \bibnamefont {Hsu}}, \bibinfo {author} {\bibfnamefont
			{F.}~\bibnamefont {Hernandez}}, \bibinfo {author} {\bibfnamefont {A.~W.}\
			\bibnamefont {Rodriguez}}, \bibinfo {author} {\bibfnamefont {M.}~\bibnamefont
			{Soljačić}},\ and\ \bibinfo {author} {\bibfnamefont {S.~G.}\ \bibnamefont
			{Johnson}},\ }\bibfield  {title} {\bibinfo {title} {General theory of
			spontaneous emission near exceptional points},\ }\href
	{https://doi.org/10.1364/OE.25.012325} {\bibfield  {journal} {\bibinfo
			{journal} {Opt. Express}\ }\textbf {\bibinfo {volume} {25}},\ \bibinfo
		{pages} {12325} (\bibinfo {year} {2017})}\BibitemShut {NoStop}%
	\bibitem [{\citenamefont {Newton}(1982)}]{Newton1982}%
	\BibitemOpen
	\bibfield  {author} {\bibinfo {author} {\bibfnamefont {R.~G.}\ \bibnamefont
			{Newton}},\ }\href {http://link.springer.com/10.1007/978-3-642-88128-2}
	{\emph {\bibinfo {title} {{Scattering Theory of Waves and Particles}}}},\
	\bibinfo {edition} {2nd}\ ed.\ (\bibinfo  {publisher} {Springer},\ \bibinfo
	{address} {Berlin},\ \bibinfo {year} {1982})\BibitemShut {NoStop}%
	\bibitem [{\citenamefont {T{\"{u}}reci}\ \emph {et~al.}(2006)\citenamefont
		{T{\"{u}}reci}, \citenamefont {Stone},\ and\ \citenamefont
		{Collier}}]{Tureci2006}%
	\BibitemOpen
	\bibfield  {author} {\bibinfo {author} {\bibfnamefont {H.~E.}\ \bibnamefont
			{T{\"{u}}reci}}, \bibinfo {author} {\bibfnamefont {A.~D.}\ \bibnamefont
			{Stone}},\ and\ \bibinfo {author} {\bibfnamefont {B.}~\bibnamefont
			{Collier}},\ }\bibfield  {title} {\bibinfo {title} {{Self-consistent
				multimode lasing theory for complex or random lasing media}},\ }\href@noop {}
	{\bibfield  {journal} {\bibinfo  {journal} {Phys. Rev. A}\ }\textbf {\bibinfo
			{volume} {74}},\ \bibinfo {pages} {043822} (\bibinfo {year}
		{2006})}\BibitemShut {NoStop}%
	\bibitem [{\citenamefont {T{\"{u}}reci}\ \emph {et~al.}(2008)\citenamefont
		{T{\"{u}}reci}, \citenamefont {Ge}, \citenamefont {Rotter},\ and\
		\citenamefont {Stone}}]{Tureci2008}%
	\BibitemOpen
	\bibfield  {author} {\bibinfo {author} {\bibfnamefont {H.~E.}\ \bibnamefont
			{T{\"{u}}reci}}, \bibinfo {author} {\bibfnamefont {L.}~\bibnamefont {Ge}},
		\bibinfo {author} {\bibfnamefont {S.}~\bibnamefont {Rotter}},\ and\ \bibinfo
		{author} {\bibfnamefont {A.~D.}\ \bibnamefont {Stone}},\ }\bibfield  {title}
	{\bibinfo {title} {{Strong interactions in multimode random lasers}},\ }\href
	{https://doi.org/10.1126/science.1155311} {\bibfield  {journal} {\bibinfo
			{journal} {Science}\ }\textbf {\bibinfo {volume} {320}},\ \bibinfo {pages}
		{643} (\bibinfo {year} {2008})}\BibitemShut {NoStop}%
	\bibitem [{\citenamefont {Ge}\ \emph {et~al.}(2010)\citenamefont {Ge},
		\citenamefont {Chong},\ and\ \citenamefont {Stone}}]{Ge2010}%
	\BibitemOpen
	\bibfield  {author} {\bibinfo {author} {\bibfnamefont {L.}~\bibnamefont
			{Ge}}, \bibinfo {author} {\bibfnamefont {Y.~D.}\ \bibnamefont {Chong}},\ and\
		\bibinfo {author} {\bibfnamefont {A.~D.}\ \bibnamefont {Stone}},\ }\bibfield
	{title} {\bibinfo {title} {{Steady-state ab initio laser theory:
				Generalizations and analytic results}},\ }\href@noop {} {\bibfield  {journal}
		{\bibinfo  {journal} {Phys. Rev. A}\ }\textbf {\bibinfo {volume} {82}},\
		\bibinfo {pages} {063824} (\bibinfo {year} {2010})}\BibitemShut {NoStop}%
	\bibitem [{\citenamefont {Qi}\ \emph {et~al.}(2018)\citenamefont {Qi},
		\citenamefont {Zhang},\ and\ \citenamefont {Ge}}]{qi_defect_2018}%
	\BibitemOpen
	\bibfield  {author} {\bibinfo {author} {\bibfnamefont {B.}~\bibnamefont
			{Qi}}, \bibinfo {author} {\bibfnamefont {L.}~\bibnamefont {Zhang}},\ and\
		\bibinfo {author} {\bibfnamefont {L.}~\bibnamefont {Ge}},\ }\bibfield
	{title} {\bibinfo {title} {Defect {States} {Emerging} from a
			{Non}-{Hermitian} {Flatband} of {Photonic} {Zero} {Modes}},\ }\href
	{https://doi.org/10.1103/PhysRevLett.120.093901} {\bibfield  {journal}
		{\bibinfo  {journal} {Phys. Rev. Lett.}\ }\textbf {\bibinfo {volume} {120}},\
		\bibinfo {pages} {093901} (\bibinfo {year} {2018})}\BibitemShut {NoStop}%
	\bibitem [{\citenamefont {Hafezi}\ \emph {et~al.}(2011)\citenamefont {Hafezi},
		\citenamefont {Demler}, \citenamefont {Lukin},\ and\ \citenamefont
		{Taylor}}]{hafezi_robust_2011}%
	\BibitemOpen
	\bibfield  {author} {\bibinfo {author} {\bibfnamefont {M.}~\bibnamefont
			{Hafezi}}, \bibinfo {author} {\bibfnamefont {E.~A.}\ \bibnamefont {Demler}},
		\bibinfo {author} {\bibfnamefont {M.~D.}\ \bibnamefont {Lukin}},\ and\
		\bibinfo {author} {\bibfnamefont {J.~M.}\ \bibnamefont {Taylor}},\ }\bibfield
	{title} {\bibinfo {title} {Robust optical delay lines with topological
			protection},\ }\href {https://doi.org/10.1038/nphys2063} {\bibfield
		{journal} {\bibinfo  {journal} {Nat. Phys.}\ }\textbf {\bibinfo {volume}
			{7}},\ \bibinfo {pages} {907} (\bibinfo {year} {2011})}\BibitemShut {NoStop}%
	\bibitem [{\citenamefont {Hafezi}\ \emph {et~al.}(2013)\citenamefont {Hafezi},
		\citenamefont {Mittal}, \citenamefont {Fan}, \citenamefont {Migdall},\ and\
		\citenamefont {Taylor}}]{hafezi_imaging_2013}%
	\BibitemOpen
	\bibfield  {author} {\bibinfo {author} {\bibfnamefont {M.}~\bibnamefont
			{Hafezi}}, \bibinfo {author} {\bibfnamefont {S.}~\bibnamefont {Mittal}},
		\bibinfo {author} {\bibfnamefont {J.}~\bibnamefont {Fan}}, \bibinfo {author}
		{\bibfnamefont {A.}~\bibnamefont {Migdall}},\ and\ \bibinfo {author}
		{\bibfnamefont {J.~M.}\ \bibnamefont {Taylor}},\ }\bibfield  {title}
	{\bibinfo {title} {Imaging topological edge states in silicon photonics},\
	}\href {https://doi.org/10.1038/nphoton.2013.274} {\bibfield  {journal}
		{\bibinfo  {journal} {Nat. Photonics}\ }\textbf {\bibinfo {volume} {7}},\
		\bibinfo {pages} {1001} (\bibinfo {year} {2013})}\BibitemShut {NoStop}%
	\bibitem [{\citenamefont {Bandres}\ \emph {et~al.}(2018)\citenamefont
		{Bandres}, \citenamefont {Wittek}, \citenamefont {Harari}, \citenamefont
		{Parto}, \citenamefont {Ren}, \citenamefont {Segev}, \citenamefont
		{Christodoulides},\ and\ \citenamefont
		{Khajavikhan}}]{bandres_topological_2018}%
	\BibitemOpen
	\bibfield  {author} {\bibinfo {author} {\bibfnamefont {M.~A.}\ \bibnamefont
			{Bandres}}, \bibinfo {author} {\bibfnamefont {S.}~\bibnamefont {Wittek}},
		\bibinfo {author} {\bibfnamefont {G.}~\bibnamefont {Harari}}, \bibinfo
		{author} {\bibfnamefont {M.}~\bibnamefont {Parto}}, \bibinfo {author}
		{\bibfnamefont {J.}~\bibnamefont {Ren}}, \bibinfo {author} {\bibfnamefont
			{M.}~\bibnamefont {Segev}}, \bibinfo {author} {\bibfnamefont {D.~N.}\
			\bibnamefont {Christodoulides}},\ and\ \bibinfo {author} {\bibfnamefont
			{M.}~\bibnamefont {Khajavikhan}},\ }\bibfield  {title} {\bibinfo {title}
		{Topological insulator laser: {Experiments}},\ }\href
	{https://doi.org/10.1126/science.aar4005} {\bibfield  {journal} {\bibinfo
			{journal} {Science}\ }\textbf {\bibinfo {volume} {359}},\ \bibinfo {pages}
		{eaar4005} (\bibinfo {year} {2018})}\BibitemShut {NoStop}%
	\bibitem [{\citenamefont {Zhao}\ \emph {et~al.}(2019)\citenamefont {Zhao},
		\citenamefont {Qiao}, \citenamefont {Wu}, \citenamefont {Midya},
		\citenamefont {Longhi},\ and\ \citenamefont
		{Feng}}]{zhao_non-hermitian_2019}%
	\BibitemOpen
	\bibfield  {author} {\bibinfo {author} {\bibfnamefont {H.}~\bibnamefont
			{Zhao}}, \bibinfo {author} {\bibfnamefont {X.}~\bibnamefont {Qiao}}, \bibinfo
		{author} {\bibfnamefont {T.}~\bibnamefont {Wu}}, \bibinfo {author}
		{\bibfnamefont {B.}~\bibnamefont {Midya}}, \bibinfo {author} {\bibfnamefont
			{S.}~\bibnamefont {Longhi}},\ and\ \bibinfo {author} {\bibfnamefont
			{L.}~\bibnamefont {Feng}},\ }\bibfield  {title} {\bibinfo {title}
		{Non-{Hermitian} topological light steering},\ }\href
	{https://doi.org/10.1126/science.aay1064} {\bibfield  {journal} {\bibinfo
			{journal} {Science}\ }\textbf {\bibinfo {volume} {365}},\ \bibinfo {pages}
		{1163} (\bibinfo {year} {2019})}\BibitemShut {NoStop}%
	\bibitem [{\citenamefont {Barik}\ \emph {et~al.}(2018)\citenamefont {Barik},
		\citenamefont {Karasahin}, \citenamefont {Flower}, \citenamefont {Cai},
		\citenamefont {Miyake}, \citenamefont {DeGottardi}, \citenamefont {Hafezi},\
		and\ \citenamefont {Waks}}]{barik_topological_2018}%
	\BibitemOpen
	\bibfield  {author} {\bibinfo {author} {\bibfnamefont {S.}~\bibnamefont
			{Barik}}, \bibinfo {author} {\bibfnamefont {A.}~\bibnamefont {Karasahin}},
		\bibinfo {author} {\bibfnamefont {C.}~\bibnamefont {Flower}}, \bibinfo
		{author} {\bibfnamefont {T.}~\bibnamefont {Cai}}, \bibinfo {author}
		{\bibfnamefont {H.}~\bibnamefont {Miyake}}, \bibinfo {author} {\bibfnamefont
			{W.}~\bibnamefont {DeGottardi}}, \bibinfo {author} {\bibfnamefont
			{M.}~\bibnamefont {Hafezi}},\ and\ \bibinfo {author} {\bibfnamefont
			{E.}~\bibnamefont {Waks}},\ }\bibfield  {title} {\bibinfo {title} {A
			topological quantum optics interface},\ }\href
	{https://doi.org/10.1126/science.aaq0327} {\bibfield  {journal} {\bibinfo
			{journal} {Science}\ }\textbf {\bibinfo {volume} {359}},\ \bibinfo {pages}
		{666} (\bibinfo {year} {2018})}\BibitemShut {NoStop}%
	\bibitem [{\citenamefont {Bender}\ and\ \citenamefont
		{Boettcher}(1998)}]{Bender1998}%
	\BibitemOpen
	\bibfield  {author} {\bibinfo {author} {\bibfnamefont {C.~M.}\ \bibnamefont
			{Bender}}\ and\ \bibinfo {author} {\bibfnamefont {S.}~\bibnamefont
			{Boettcher}},\ }\bibfield  {title} {\bibinfo {title} {Real spectra in
			non-{Hermitian} {Hamiltonians} having {PT} symmetry},\ }\href
	{http://link.aps.org/doi/10.1103/PhysRevLett.80.5243} {\bibfield  {journal}
		{\bibinfo  {journal} {Phys. Rev. Lett.}\ }\textbf {\bibinfo {volume} {80}},\
		\bibinfo {pages} {5243} (\bibinfo {year} {1998})}\BibitemShut {NoStop}%
	\bibitem [{\citenamefont {Feng}\ \emph {et~al.}(2017)\citenamefont {Feng},
		\citenamefont {El-Ganainy},\ and\ \citenamefont {Ge}}]{Feng2017}%
	\BibitemOpen
	\bibfield  {author} {\bibinfo {author} {\bibfnamefont {L.}~\bibnamefont
			{Feng}}, \bibinfo {author} {\bibfnamefont {R.}~\bibnamefont {El-Ganainy}},\
		and\ \bibinfo {author} {\bibfnamefont {L.}~\bibnamefont {Ge}},\ }\bibfield
	{title} {\bibinfo {title} {Non-{Hermitian} photonics based on parity–time
			symmetry},\ }\href {https://doi.org/10.1038/s41566-017-0031-1} {\bibfield
		{journal} {\bibinfo  {journal} {Nat. Photonics}\ }\textbf {\bibinfo {volume}
			{11}},\ \bibinfo {pages} {752} (\bibinfo {year} {2017})}\BibitemShut
	{NoStop}%
	\bibitem [{\citenamefont {Chen}\ \emph {et~al.}(2020)\citenamefont {Chen},
		\citenamefont {Liu}, \citenamefont {Luan}, \citenamefont {Liu}, \citenamefont
		{Wang}, \citenamefont {Zhu}, \citenamefont {Li}, \citenamefont {Gu},
		\citenamefont {Liang}, \citenamefont {Gao}, \citenamefont {Lu}, \citenamefont
		{Ge}, \citenamefont {Zhang}, \citenamefont {Zhu},\ and\ \citenamefont
		{Ma}}]{Chen2020}%
	\BibitemOpen
	\bibfield  {author} {\bibinfo {author} {\bibfnamefont {H.~Z.}\ \bibnamefont
			{Chen}}, \bibinfo {author} {\bibfnamefont {T.}~\bibnamefont {Liu}}, \bibinfo
		{author} {\bibfnamefont {H.~Y.}\ \bibnamefont {Luan}}, \bibinfo {author}
		{\bibfnamefont {R.~J.}\ \bibnamefont {Liu}}, \bibinfo {author} {\bibfnamefont
			{X.~Y.}\ \bibnamefont {Wang}}, \bibinfo {author} {\bibfnamefont {X.~F.}\
			\bibnamefont {Zhu}}, \bibinfo {author} {\bibfnamefont {Y.~B.}\ \bibnamefont
			{Li}}, \bibinfo {author} {\bibfnamefont {Z.~M.}\ \bibnamefont {Gu}}, \bibinfo
		{author} {\bibfnamefont {S.~J.}\ \bibnamefont {Liang}}, \bibinfo {author}
		{\bibfnamefont {H.}~\bibnamefont {Gao}}, \bibinfo {author} {\bibfnamefont
			{L.}~\bibnamefont {Lu}}, \bibinfo {author} {\bibfnamefont {L.}~\bibnamefont
			{Ge}}, \bibinfo {author} {\bibfnamefont {S.}~\bibnamefont {Zhang}}, \bibinfo
		{author} {\bibfnamefont {J.}~\bibnamefont {Zhu}},\ and\ \bibinfo {author}
		{\bibfnamefont {R.~M.}\ \bibnamefont {Ma}},\ }\bibfield  {title} {\bibinfo
		{title} {Revealing the missing dimension at an exceptional point},\
	}\href@noop {} {\bibfield  {journal} {\bibinfo  {journal} {Nat. Phys.}\
		}\textbf {\bibinfo {volume} {16}},\ \bibinfo {pages} {571} (\bibinfo {year}
		{2020})}\BibitemShut {NoStop}%
	\bibitem [{\citenamefont {Beenakker}(1997)}]{Beenakker1997}%
	\BibitemOpen
	\bibfield  {author} {\bibinfo {author} {\bibfnamefont {C.~W.~J.}\
			\bibnamefont {Beenakker}},\ }\bibfield  {title} {\bibinfo {title}
		{Random-matrix theory of quantum transport},\ }\href
	{https://doi.org/10.1103/RevModPhys.69.731} {\bibfield  {journal} {\bibinfo
			{journal} {Rev. Mod. Phys.}\ }\textbf {\bibinfo {volume} {69}},\ \bibinfo
		{pages} {731} (\bibinfo {year} {1997})}\BibitemShut {NoStop}%
	\bibitem [{\citenamefont {Imry}(2008)}]{Imry2008}%
	\BibitemOpen
	\bibfield  {author} {\bibinfo {author} {\bibfnamefont {Y.}~\bibnamefont
			{Imry}},\ }\href@noop {} {\emph {\bibinfo {title} {Introduction to
				{Mesoscopic} {Physics}}}},\ \bibinfo {edition} {2nd}\ ed.\ (\bibinfo
	{publisher} {Oxford University Press},\ \bibinfo {address} {Oxford},\
	\bibinfo {year} {2008})\BibitemShut {NoStop}%
	\bibitem [{\citenamefont {Mello}\ and\ \citenamefont
		{Kumar}(2004)}]{Mello2004}%
	\BibitemOpen
	\bibfield  {author} {\bibinfo {author} {\bibfnamefont {P.~A.}\ \bibnamefont
			{Mello}}\ and\ \bibinfo {author} {\bibfnamefont {N.}~\bibnamefont {Kumar}},\
	}\href {https://doi.org/10.1093/acprof:oso/9780198525820.001.0001} {\emph
		{\bibinfo {title} {{Quantum Transport in Mesoscopic Systems}}}}\ (\bibinfo
	{publisher} {Oxford University Press},\ \bibinfo {year} {2004})\BibitemShut
	{NoStop}%
	\bibitem [{\citenamefont {Ge}(2017{\natexlab{a}})}]{Ge2017}%
	\BibitemOpen
	\bibfield  {author} {\bibinfo {author} {\bibfnamefont {L.}~\bibnamefont
			{Ge}},\ }\bibfield  {title} {\bibinfo {title} {{Constructing the scattering
				matrix for optical microcavities as a nonlocal boundary value problem}},\
	}\href {https://doi.org/10.1364/prj.5.000b20} {\bibfield  {journal} {\bibinfo
			{journal} {Photonics Res.}\ }\textbf {\bibinfo {volume} {5}},\ \bibinfo
		{pages} {B20} (\bibinfo {year} {2017}{\natexlab{a}})}\BibitemShut {NoStop}%
	\bibitem [{\citenamefont {Liertzer}\ \emph {et~al.}(2012)\citenamefont
		{Liertzer}, \citenamefont {Ge}, \citenamefont {Cerjan}, \citenamefont
		{Stone}, \citenamefont {T{\"{u}}reci},\ and\ \citenamefont
		{Rotter}}]{Liertzer2012}%
	\BibitemOpen
	\bibfield  {author} {\bibinfo {author} {\bibfnamefont {M.}~\bibnamefont
			{Liertzer}}, \bibinfo {author} {\bibfnamefont {L.}~\bibnamefont {Ge}},
		\bibinfo {author} {\bibfnamefont {A.}~\bibnamefont {Cerjan}}, \bibinfo
		{author} {\bibfnamefont {A.~D.}\ \bibnamefont {Stone}}, \bibinfo {author}
		{\bibfnamefont {H.~E.}\ \bibnamefont {T{\"{u}}reci}},\ and\ \bibinfo {author}
		{\bibfnamefont {S.}~\bibnamefont {Rotter}},\ }\bibfield  {title} {\bibinfo
		{title} {{Pump-induced exceptional points in lasers}},\ }\href@noop {}
	{\bibfield  {journal} {\bibinfo  {journal} {Phys. Rev. Lett.}\ }\textbf
		{\bibinfo {volume} {108}},\ \bibinfo {pages} {173901} (\bibinfo {year}
		{2012})}\BibitemShut {NoStop}%
	\bibitem [{\citenamefont {Heiss}(2004)}]{Heiss2004}%
	\BibitemOpen
	\bibfield  {author} {\bibinfo {author} {\bibfnamefont {W.~D.}\ \bibnamefont
			{Heiss}},\ }\bibfield  {title} {\bibinfo {title} {Exceptional points of
			non-{Hermitian} operators},\ }\href
	{http://iopscience.iop.org/0305-4470/37/6/034} {\bibfield  {journal}
		{\bibinfo  {journal} {J. Phys. A}\ }\textbf {\bibinfo {volume} {37}},\
		\bibinfo {pages} {2455} (\bibinfo {year} {2004})}\BibitemShut {NoStop}%
	\bibitem [{\citenamefont {Berry}(2004)}]{Berry2004}%
	\BibitemOpen
	\bibfield  {author} {\bibinfo {author} {\bibfnamefont {M.~V.}\ \bibnamefont
			{Berry}},\ }\bibfield  {title} {\bibinfo {title} {{Physics of nonhermitian
				degeneracies}},\ }\href {https://doi.org/10.1023/B:CJOP.0000044002.05657.04}
	{\bibfield  {journal} {\bibinfo  {journal} {Czech. J. Phys.}\ }\textbf
		{\bibinfo {volume} {54}},\ \bibinfo {pages} {1039} (\bibinfo {year}
		{2004})}\BibitemShut {NoStop}%
	\bibitem [{\citenamefont {Ge}\ and\ \citenamefont {El-Ganainy}(2016)}]{Ge2016}%
	\BibitemOpen
	\bibfield  {author} {\bibinfo {author} {\bibfnamefont {L.}~\bibnamefont
			{Ge}}\ and\ \bibinfo {author} {\bibfnamefont {R.}~\bibnamefont
			{El-Ganainy}},\ }\bibfield  {title} {\bibinfo {title} {{Nonlinear modal
				interactions in parity-time (PT) symmetric lasers}},\ }\href@noop {}
	{\bibfield  {journal} {\bibinfo  {journal} {Sci. Rep.}\ }\textbf {\bibinfo
			{volume} {6}},\ \bibinfo {pages} {24889} (\bibinfo {year}
		{2016})}\BibitemShut {NoStop}%
	\bibitem [{\citenamefont {Kapur}\ and\ \citenamefont
		{Peierls}(1937)}]{Kapur1937}%
	\BibitemOpen
	\bibfield  {author} {\bibinfo {author} {\bibfnamefont {P.~L.}\ \bibnamefont
			{Kapur}}\ and\ \bibinfo {author} {\bibfnamefont {R.}~\bibnamefont
			{Peierls}},\ }\bibfield  {title} {\bibinfo {title} {{The dispersion formula
				for nuclear reactions}},\ }\href@noop {} {\bibfield  {journal} {\bibinfo
			{journal} {P. Roy Soc. A-Math. Phy.}\ }\textbf {\bibinfo {volume} {166}},\
		\bibinfo {pages} {277} (\bibinfo {year} {1937})}\BibitemShut {NoStop}%
	\bibitem [{\citenamefont {Goldberger}\ and\ \citenamefont
		{Watson}(1964)}]{Goldberger1964}%
	\BibitemOpen
	\bibfield  {author} {\bibinfo {author} {\bibfnamefont {M.~L.}\ \bibnamefont
			{Goldberger}}\ and\ \bibinfo {author} {\bibfnamefont {K.~M.}\ \bibnamefont
			{Watson}},\ }\bibfield  {title} {\bibinfo {title} {{Lifetime and decay of
				unstable particles in S-matrix theory}},\ }\href@noop {} {\bibfield
		{journal} {\bibinfo  {journal} {Phys. Rev.}\ }\textbf {\bibinfo {volume}
			{136}},\ \bibinfo {pages} {B1472} (\bibinfo {year} {1964})}\BibitemShut
	{NoStop}%
	\bibitem [{\citenamefont {Heiss}(2015)}]{Heiss2015}%
	\BibitemOpen
	\bibfield  {author} {\bibinfo {author} {\bibfnamefont {W.~D.}\ \bibnamefont
			{Heiss}},\ }\bibfield  {title} {\bibinfo {title} {{Green's Functions at
				Exceptional Points}},\ }\href {https://doi.org/10.1007/s10773-014-2428-7}
	{\bibfield  {journal} {\bibinfo  {journal} {Int. J. Theor. Phys.}\ }\textbf
		{\bibinfo {volume} {54}},\ \bibinfo {pages} {3954} (\bibinfo {year}
		{2015})}\BibitemShut {NoStop}%
	\bibitem [{\citenamefont {Guo}\ \emph {et~al.}(2009)\citenamefont {Guo},
		\citenamefont {Salamo}, \citenamefont {Duchesne}, \citenamefont {Morandotti},
		\citenamefont {Volatier-Ravat}, \citenamefont {Aimez}, \citenamefont
		{Siviloglou},\ and\ \citenamefont {Christodoulides}}]{Guo}%
	\BibitemOpen
	\bibfield  {author} {\bibinfo {author} {\bibfnamefont {A.}~\bibnamefont
			{Guo}}, \bibinfo {author} {\bibfnamefont {G.~J.}\ \bibnamefont {Salamo}},
		\bibinfo {author} {\bibfnamefont {D.}~\bibnamefont {Duchesne}}, \bibinfo
		{author} {\bibfnamefont {R.}~\bibnamefont {Morandotti}}, \bibinfo {author}
		{\bibfnamefont {M.}~\bibnamefont {Volatier-Ravat}}, \bibinfo {author}
		{\bibfnamefont {V.}~\bibnamefont {Aimez}}, \bibinfo {author} {\bibfnamefont
			{G.~A.}\ \bibnamefont {Siviloglou}},\ and\ \bibinfo {author} {\bibfnamefont
			{D.~N.}\ \bibnamefont {Christodoulides}},\ }\bibfield  {title} {\bibinfo
		{title} {Observation of {PT}-symmetry breaking in complex optical
			potentials},\ }\href {http://link.aps.org/doi/10.1103/PhysRevLett.103.093902}
	{\bibfield  {journal} {\bibinfo  {journal} {Phys. Rev. Lett.}\ }\textbf
		{\bibinfo {volume} {103}},\ \bibinfo {pages} {93902} (\bibinfo {year}
		{2009})}\BibitemShut {NoStop}%
	\bibitem [{\citenamefont {Miao}\ \emph {et~al.}(2016)\citenamefont {Miao},
		\citenamefont {Zhang}, \citenamefont {Sun}, \citenamefont {Walasik},
		\citenamefont {Longhi}, \citenamefont {Litchinitser},\ and\ \citenamefont
		{Feng}}]{Miao2016}%
	\BibitemOpen
	\bibfield  {author} {\bibinfo {author} {\bibfnamefont {P.}~\bibnamefont
			{Miao}}, \bibinfo {author} {\bibfnamefont {Z.}~\bibnamefont {Zhang}},
		\bibinfo {author} {\bibfnamefont {J.}~\bibnamefont {Sun}}, \bibinfo {author}
		{\bibfnamefont {W.}~\bibnamefont {Walasik}}, \bibinfo {author} {\bibfnamefont
			{S.}~\bibnamefont {Longhi}}, \bibinfo {author} {\bibfnamefont {N.~M.}\
			\bibnamefont {Litchinitser}},\ and\ \bibinfo {author} {\bibfnamefont
			{L.}~\bibnamefont {Feng}},\ }\bibfield  {title} {\bibinfo {title} {Orbital
			angular momentum microlaser},\ }\href
	{http://science.sciencemag.org/content/353/6298/464.short} {\bibfield
		{journal} {\bibinfo  {journal} {Science}\ }\textbf {\bibinfo {volume}
			{353}},\ \bibinfo {pages} {464} (\bibinfo {year} {2016})}\BibitemShut
	{NoStop}%
	\bibitem [{\citenamefont {Fisher}\ and\ \citenamefont
		{Lee}(1981)}]{fisher_relation_1981}%
	\BibitemOpen
	\bibfield  {author} {\bibinfo {author} {\bibfnamefont {D.~S.}\ \bibnamefont
			{Fisher}}\ and\ \bibinfo {author} {\bibfnamefont {P.~A.}\ \bibnamefont
			{Lee}},\ }\bibfield  {title} {\bibinfo {title} {Relation between conductivity
			and transmission matrix},\ }\href
	{https://link.aps.org/doi/10.1103/PhysRevB.23.6851} {\bibfield  {journal}
		{\bibinfo  {journal} {Phys. Rev. B}\ }\textbf {\bibinfo {volume} {23}},\
		\bibinfo {pages} {6851} (\bibinfo {year} {1981})}\BibitemShut {NoStop}%
	\bibitem [{\citenamefont {Stone}\ and\ \citenamefont
		{Szafer}(1988)}]{stone_what_1988}%
	\BibitemOpen
	\bibfield  {author} {\bibinfo {author} {\bibfnamefont {A.~D.}\ \bibnamefont
			{Stone}}\ and\ \bibinfo {author} {\bibfnamefont {A.}~\bibnamefont {Szafer}},\
	}\bibfield  {title} {\bibinfo {title} {What is measured when you measure a
			resistance? {T}he landauer formula revisited},\ }\href@noop {} {\bibfield
		{journal} {\bibinfo  {journal} {IBM J. Res. Dev.}\ }\textbf {\bibinfo
			{volume} {32}},\ \bibinfo {pages} {384} (\bibinfo {year} {1988})}\BibitemShut
	{NoStop}%
	\bibitem [{\citenamefont {Ge}\ \emph {et~al.}(2015)\citenamefont {Ge},
		\citenamefont {Makris}, \citenamefont {Christodoulides},\ and\ \citenamefont
		{Feng}}]{Ge_scattering_2015}%
	\BibitemOpen
	\bibfield  {author} {\bibinfo {author} {\bibfnamefont {L.}~\bibnamefont
			{Ge}}, \bibinfo {author} {\bibfnamefont {K.~G.}\ \bibnamefont {Makris}},
		\bibinfo {author} {\bibfnamefont {D.~N.}\ \bibnamefont {Christodoulides}},\
		and\ \bibinfo {author} {\bibfnamefont {L.}~\bibnamefont {Feng}},\ }\bibfield
	{title} {\bibinfo {title} {Scattering in {PT}- and {RT}-symmetric multimode
			waveguides: {Generalized} conservation laws and spontaneous symmetry breaking
			beyond one dimension},\ }\href {https://doi.org/10.1103/PhysRevA.92.062135}
	{\bibfield  {journal} {\bibinfo  {journal} {Phys. Rev. A}\ }\textbf {\bibinfo
			{volume} {92}},\ \bibinfo {pages} {062135} (\bibinfo {year}
		{2015})}\BibitemShut {NoStop}%
	\bibitem [{\citenamefont {Thouless}\ and\ \citenamefont
		{Kirkpatrick}(1981)}]{thouless_conductivity_1981}%
	\BibitemOpen
	\bibfield  {author} {\bibinfo {author} {\bibfnamefont {D.~J.}\ \bibnamefont
			{Thouless}}\ and\ \bibinfo {author} {\bibfnamefont {S.}~\bibnamefont
			{Kirkpatrick}},\ }\bibfield  {title} {\bibinfo {title} {Conductivity of the
			disordered linear chain},\ }\href
	{https://doi.org/10.1088/0022-3719/14/3/007} {\bibfield  {journal} {\bibinfo
			{journal} {J. Phys. C}\ }\textbf {\bibinfo {volume} {14}},\ \bibinfo {pages}
		{235} (\bibinfo {year} {1981})}\BibitemShut {NoStop}%
	\bibitem [{\citenamefont {Sols}\ \emph {et~al.}(1989)\citenamefont {Sols},
		\citenamefont {Macucci}, \citenamefont {Ravaioli},\ and\ \citenamefont
		{Hess}}]{Sols_theory_1989}%
	\BibitemOpen
	\bibfield  {author} {\bibinfo {author} {\bibfnamefont {F.}~\bibnamefont
			{Sols}}, \bibinfo {author} {\bibfnamefont {M.}~\bibnamefont {Macucci}},
		\bibinfo {author} {\bibfnamefont {U.}~\bibnamefont {Ravaioli}},\ and\
		\bibinfo {author} {\bibfnamefont {K.}~\bibnamefont {Hess}},\ }\bibfield
	{title} {\bibinfo {title} {Theory for a quantum modulated transistor},\
	}\href {https://doi.org/10.1063/1.344032} {\bibfield  {journal} {\bibinfo
			{journal} {J. Appl. Phys.}\ }\textbf {\bibinfo {volume} {66}},\ \bibinfo
		{pages} {3892} (\bibinfo {year} {1989})}\BibitemShut {NoStop}%
	\bibitem [{\citenamefont {Patil}(2006)}]{Patil2006}%
	\BibitemOpen
	\bibfield  {author} {\bibinfo {author} {\bibfnamefont {S.~H.}\ \bibnamefont
			{Patil}},\ }\bibfield  {title} {\bibinfo {title} {Harmonic oscillator with a
			\${\textbackslash}updelta\$-function potential},\ }\href
	{https://doi.org/10.1088/0143-0807/27/4/021} {\bibfield  {journal} {\bibinfo
			{journal} {Eur. J. Phys.}\ }\textbf {\bibinfo {volume} {27}},\ \bibinfo
		{pages} {899} (\bibinfo {year} {2006})}\BibitemShut {NoStop}%
	\bibitem [{\citenamefont {Malzard}\ \emph {et~al.}(2015)\citenamefont
		{Malzard}, \citenamefont {Poli},\ and\ \citenamefont {Schomerus}}]{Malzard}%
	\BibitemOpen
	\bibfield  {author} {\bibinfo {author} {\bibfnamefont {S.}~\bibnamefont
			{Malzard}}, \bibinfo {author} {\bibfnamefont {C.}~\bibnamefont {Poli}},\ and\
		\bibinfo {author} {\bibfnamefont {H.}~\bibnamefont {Schomerus}},\ }\bibfield
	{title} {\bibinfo {title} {Topologically {Protected} {Defect} {States} in
			{Open} {Photonic} {Systems} with {Non}-{Hermitian} {Charge}-{Conjugation} and
			{Parity}-{Time} {Symmetry}},\ }\href
	{https://doi.org/10.1103/PhysRevLett.115.200402} {\bibfield  {journal}
		{\bibinfo  {journal} {Phys. Rev. Lett.}\ }\textbf {\bibinfo {volume} {115}},\
		\bibinfo {pages} {200402} (\bibinfo {year} {2015})}\BibitemShut {NoStop}%
	\bibitem [{\citenamefont {Ge}(2017{\natexlab{b}})}]{zeromodeLaser}%
	\BibitemOpen
	\bibfield  {author} {\bibinfo {author} {\bibfnamefont {L.}~\bibnamefont
			{Ge}},\ }\bibfield  {title} {\bibinfo {title} {Symmetry-protected zero-mode
			laser with a tunable spatial profile},\ }\href
	{https://doi.org/10.1103/PhysRevA.95.023812} {\bibfield  {journal} {\bibinfo
			{journal} {Phys. Rev. A}\ }\textbf {\bibinfo {volume} {95}},\ \bibinfo
		{pages} {023812} (\bibinfo {year} {2017}{\natexlab{b}})}\BibitemShut
	{NoStop}%
	\bibitem [{\citenamefont {Rivero}\ and\ \citenamefont {Ge}(2021)}]{NHC_arxiv}%
	\BibitemOpen
	\bibfield  {author} {\bibinfo {author} {\bibfnamefont {J.~D.~H.}\
			\bibnamefont {Rivero}}\ and\ \bibinfo {author} {\bibfnamefont
			{L.}~\bibnamefont {Ge}},\ }\bibfield  {title} {\bibinfo {title} {Chiral
			symmetry in non-hermitian systems: Product rule and clifford algebra},\
	}\href@noop {} {\bibfield  {journal} {\bibinfo  {journal} {Phys. Rev. B}\
		}\textbf {\bibinfo {volume} {103}},\ \bibinfo {pages} {014111} (\bibinfo
		{year} {2021})}\BibitemShut {NoStop}%
	\bibitem [{\citenamefont {Kawabata}\ \emph {et~al.}(2019)\citenamefont
		{Kawabata}, \citenamefont {Higashikawa}, \citenamefont {Gong}, \citenamefont
		{Ashida},\ and\ \citenamefont {Ueda}}]{Kawabata}%
	\BibitemOpen
	\bibfield  {author} {\bibinfo {author} {\bibfnamefont {K.}~\bibnamefont
			{Kawabata}}, \bibinfo {author} {\bibfnamefont {S.}~\bibnamefont
			{Higashikawa}}, \bibinfo {author} {\bibfnamefont {Z.}~\bibnamefont {Gong}},
		\bibinfo {author} {\bibfnamefont {Y.}~\bibnamefont {Ashida}},\ and\ \bibinfo
		{author} {\bibfnamefont {M.}~\bibnamefont {Ueda}},\ }\bibfield  {title}
	{\bibinfo {title} {Topological unification of time-reversal and particle-hole
			symmetries in non-{Hermitian} physics},\ }\href
	{https://doi.org/10.1038/s41467-018-08254-y} {\bibfield  {journal} {\bibinfo
			{journal} {Nature Commun.}\ }\textbf {\bibinfo {volume} {10}},\ \bibinfo
		{pages} {297} (\bibinfo {year} {2019})}\BibitemShut {NoStop}%
	\bibitem [{\citenamefont {Baboux}\ \emph {et~al.}(2016)\citenamefont {Baboux},
		\citenamefont {Ge}, \citenamefont {Jacqmin}, \citenamefont {Biondi},
		\citenamefont {Galopin}, \citenamefont {Lemaître}, \citenamefont
		{Le~Gratiet}, \citenamefont {Sagnes}, \citenamefont {Schmidt}, \citenamefont
		{Türeci}, \citenamefont {Amo},\ and\ \citenamefont {Bloch}}]{Baboux}%
	\BibitemOpen
	\bibfield  {author} {\bibinfo {author} {\bibfnamefont {F.}~\bibnamefont
			{Baboux}}, \bibinfo {author} {\bibfnamefont {L.}~\bibnamefont {Ge}}, \bibinfo
		{author} {\bibfnamefont {T.}~\bibnamefont {Jacqmin}}, \bibinfo {author}
		{\bibfnamefont {M.}~\bibnamefont {Biondi}}, \bibinfo {author} {\bibfnamefont
			{E.}~\bibnamefont {Galopin}}, \bibinfo {author} {\bibfnamefont
			{A.}~\bibnamefont {Lemaitre}}, \bibinfo {author} {\bibfnamefont
			{L.}~\bibnamefont {Le~Gratiet}}, \bibinfo {author} {\bibfnamefont
			{I.}~\bibnamefont {Sagnes}}, \bibinfo {author} {\bibfnamefont
			{S.}~\bibnamefont {Schmidt}}, \bibinfo {author} {\bibfnamefont
			{H.~E.}~\bibnamefont {Tureci}}, \bibinfo {author} {\bibfnamefont
			{A.}~\bibnamefont {Amo}},\ and\ \bibinfo {author} {\bibfnamefont
			{J.}~\bibnamefont {Bloch}},\ }\bibfield  {title} {\bibinfo {title} {Bosonic
			{Condensation} and {Disorder}-{Induced} {Localization} in a {Flat} {Band}},\
	}\href {https://doi.org/10.1103/PhysRevLett.116.066402} {\bibfield  {journal}
		{\bibinfo  {journal} {Phys. Rev. Lett.}\ }\textbf {\bibinfo {volume} {116}},\
		\bibinfo {pages} {066402} (\bibinfo {year} {2016})}\BibitemShut {NoStop}%
	\bibitem [{\citenamefont {Landau}\ and\ \citenamefont
		{Lifshitz}(1960)}]{Landau}%
	\BibitemOpen
	\bibfield  {author} {\bibinfo {author} {\bibfnamefont {L.~D.}\ \bibnamefont
			{Landau}}\ and\ \bibinfo {author} {\bibfnamefont {E.~M.}\ \bibnamefont
			{Lifshitz}},\ }\href@noop {} {\emph {\bibinfo {title} {Electrodynamics of
				Continuous Media}}}\ (\bibinfo  {publisher} {Pergamon Press},\ \bibinfo
	{address} {Oxford},\ \bibinfo {year} {1960})\BibitemShut {NoStop}%
	\bibitem [{\citenamefont {Ge}\ and\ \citenamefont
		{Feng}(2016)}]{ge_reciprocity2016}%
	\BibitemOpen
	\bibfield  {author} {\bibinfo {author} {\bibfnamefont {L.}~\bibnamefont
			{Ge}}\ and\ \bibinfo {author} {\bibfnamefont {L.}~\bibnamefont {Feng}},\
	}\bibfield  {title} {\bibinfo {title} {Optical-reciprocity-induced symmetry
			in photonic heterostructures and its manifestation in scattering
			{PT}-symmetry breaking},\ }\href {https://doi.org/10.1103/PhysRevA.94.043836}
	{\bibfield  {journal} {\bibinfo  {journal} {Phys. Rev. A}\ }\textbf {\bibinfo
			{volume} {94}},\ \bibinfo {pages} {043836} (\bibinfo {year}
		{2016})}\BibitemShut {NoStop}%
	\bibitem [{\citenamefont {Hasan}\ and\ \citenamefont {Kane}(2010)}]{Hasan}%
	\BibitemOpen
	\bibfield  {author} {\bibinfo {author} {\bibfnamefont {M.~Z.}\ \bibnamefont
			{Hasan}}\ and\ \bibinfo {author} {\bibfnamefont {C.~L.}\ \bibnamefont
			{Kane}},\ }\bibfield  {title} {\bibinfo {title} {Colloquium: {Topological}
			insulators},\ }\href {https://doi.org/10.1103/RevModPhys.82.3045} {\bibfield
		{journal} {\bibinfo  {journal} {Rev. Mod. Phys.}\ }\textbf {\bibinfo {volume}
			{82}},\ \bibinfo {pages} {3045} (\bibinfo {year} {2010})}\BibitemShut
	{NoStop}%
	\bibitem [{\citenamefont {Gottfried}\ and\ \citenamefont
		{Yan}(2003)}]{Gottfried}%
	\BibitemOpen
	\bibfield  {author} {\bibinfo {author} {\bibfnamefont {K.}~\bibnamefont
			{Gottfried}}\ and\ \bibinfo {author} {\bibfnamefont {T.-M.}\ \bibnamefont
			{Yan}},\ }\href@noop {} {\emph {\bibinfo {title} {{Q}uantum {M}echanics:
				{F}undamentals}}},\ \bibinfo {edition} {2nd}\ ed.\ (\bibinfo  {publisher}
	{Springer},\ \bibinfo {address} {New York},\ \bibinfo {year}
	{2003})\BibitemShut {NoStop}%
	\bibitem [{\citenamefont {Rivero}\ and\ \citenamefont
		{Ge}(2020)}]{Noether_nonHermitian}%
	\BibitemOpen
	\bibfield  {author} {\bibinfo {author} {\bibfnamefont {J.~D.~H.}\ \bibnamefont
			{Rivero}}\ and\ \bibinfo {author} {\bibfnamefont {L.}~\bibnamefont {Ge}},\
	}\bibfield  {title} {\bibinfo {title} {Pseudochirality: {A} {Manifestation}
			of {Noether}'s {Theorem} in {Non}-{Hermitian} {Systems}},\ }\href@noop {}
	{\bibfield  {journal} {\bibinfo  {journal} {Phys. Rev. Lett.}\ }\textbf
		{\bibinfo {volume} {125}},\ \bibinfo {pages} {083902} (\bibinfo {year}
		{2020})}\BibitemShut {NoStop}%
	\bibitem [{\citenamefont {Rivero}\ and\ \citenamefont
		{Ge}(2019)}]{Jose_scaling2019}%
	\BibitemOpen
	\bibfield  {author} {\bibinfo {author} {\bibfnamefont {J.~D.~H.}\
			\bibnamefont {Rivero}}\ and\ \bibinfo {author} {\bibfnamefont
			{L.}~\bibnamefont {Ge}},\ }\bibfield  {title} {\bibinfo {title}
		{Time-reversal-invariant scaling of light propagation in one-dimensional
			non-hermitian systems},\ }\href {https://doi.org/10.1103/PhysRevA.100.023819}
	{\bibfield  {journal} {\bibinfo  {journal} {Phys. Rev. A}\ }\textbf {\bibinfo
			{volume} {100}},\ \bibinfo {pages} {023819} (\bibinfo {year}
		{2019})}\BibitemShut {NoStop}%
\end{thebibliography}
\end{document}